\documentclass{pasa}%

\title[Evolutionary Models for SN Ib/c Progenitors]{Evolutionary Models for Type Ib/c Supernova Progenitors}
\author[Sung-Chul Yoon]{Sung-Chul Yoon$^1$ \\
\affil{$^1$Department of Physics and Astronomy, Seoul National University, Gwanak-ro 1, Gwanak-gu, Seoul, 151-742, Republic of Korea; yoon@astro.snu.ac.kr}}%
\jid{PASA}
\doi{10.1017/pas.\the\year.xxx}
\jyear{\the\year}

\usepackage[authoryear]{natbib}
\usepackage{lscape}
\usepackage{pdflscape}
\usepackage{rotating}
\usepackage{lipsum}
\usepackage{amsmath}
\usepackage{tabularx}
\usepackage{threeparttable}
\usepackage{float}
\usepackage{graphicx}

\bibpunct{(}{)}{;}{a}{}{,}
\newcommand{\Msun}{${\mathrm{M_\odot}}$}

\begin{document}
\begin{abstract}
Type Ib/c  supernovae (SNe Ib/c) mark the deaths of hydrogen-deficient massive
stars.  The evolutionary scenarios for SNe Ib/c progenitors involve many important physical
processes including mass loss by winds and its metallicity dependence, stellar
rotation, and binary interactions. This makes SNe Ib/c an excellent test bed
for stellar evolution theory.  We review the main results of evolutionary
models for SN Ib/c progenitors available in the literature and their
confrontation with recent observations. We argue that the nature of SN Ib/c
progenitors can be significantly different for single and binary systems, and
that binary evolution models can explain the ejecta masses derived from SN Ib/c
light curves,  the distribution of SN Ib/c sites in their host galaxies, and
the optical magnitudes of the tentative progenitor candidate of iPTF13bvn.  We
emphasize the importance of early-time observations of light curves and
spectra,  accurate measurements of helium mass in SN Ib/c ejecta, and systematic
studies about the metallicity dependence of SN Ib/c properties, to better
constrain theories.  
\end{abstract}
\begin{keywords} stars:evolution -- binaries:general -- supernovae:general --
stars: massive -- stars: Wolf-Rayet \end{keywords}
\maketitle%
\section{INTRODUCTION } \label{sec:intro}

Type I supernovae (SNe I) are characterized by the lack of prominent hydrogen lines in
the spectra \citep[e.g.,][]{Filippenko97}.  Strong helium lines are present in
the spectra of SNe Ib, while they are practically absent
in those of SNe Ic.   SNe Ib/c are further distinguished
from SNe Ia by the lack of strong SiII absorption line at
6355~\AA.  Most of ordinary SNe Ib/c, if not all, occur in star-forming
galaxies, indicating that SNe Ib/c have a massive star origin
\citep[e.g.][]{Berg05, Boissier09, Hakobyan09, Kelly12, Anderson12, Sanders12}.
Their light curves are dominated by the energy release from radioactive
$^{56}$Ni as in the case of SNe Ia \citep{Schaeffer87}, but  the inferred
amounts of $^{56}$Ni ejected by SNe Ib/c are similar, on average, to those of
SNe II ($M_\mathrm{^{56}Ni}\sim 0.1~\mathrm{M_\odot}$;e.g., \citealt{Drout11,
Cano13, Lyman14, Taddia14}) rather than  SNe Ia ($M_\mathrm{^{56}Ni}\sim
1.0~\mathrm{M_\odot}$; e.g. \citealt{Stritzinger06, Mazzali07, Scalzo14}).  The
current consensus is that most of SNe Ib/c belong to a subset of core-collapse
SNe that are SN explosions via collapse of the iron cores in massive stars at
their deaths. 

Hydrogen cannot be easily hidden in  SN spectra \citep[e.g.,][]{Elmhamdi06,
Spencer10, Dessart11, Hachinger12} and  SN Ib/c progenitors must have lost
their hydrogen envelopes by the time of explosion.  There exist mainly three
possible ways for massive stars to become a hydrogen-deficient SN progenitor:
mass loss from single stars via stellar winds \citep[e.g.,][]{Chiosi86}, binary
interactions \citep[e.g.][]{Podsiadlowski92}, and chemically homogeneous
evolution with rapid rotation \citep{Maeder87}. The last mode of evolution have
been invoked for explaining massive blue stragglers and long gamma-ray bursts
within the collapsar scenario \citep{Maeder87a, Langer92, Yoon05, Woosley06,
Yoon06, Yoon12a}, but is not likely to be much relevant for the majority of SNe
Ib/c that are found in the local Universe \citep{Yoon06}.  

In this review, we focus on ordinary SNe Ib/c: our objective here is to
summarize theoretical results on  SN Ib/c progenitors via single and binary
evolutionary paths.   We emphasize that each case has its own unique
prediction that can be in principle well tested by observations. SNe Ib/c
can therefore provide an invaluable insight on massive star evolution.  Note
that we restrict this review  to the detailed properties of SNe Ib/c
progenitors that are predicted by recent stellar evolution models.  Progenitors
of SN IIb (i.e., SNe of which the spectra have hydrogen lines at early
times, but resemble those of SNe Ib at later times) are closely related to SN
Ib/c progenitors, and  will also be discussed briefly.  Our discussions on
observations,  SN modeling, and stellar population studies will be
highly biased by the selected topics we address here.  For more general topics
on the evolution of massive stars and SN progenitors, readers are
referred to the recent reviews by \citet{Maeder00},  \citet{Massey03},
\citet{Heger03}, \citet{Smartt09}, \citet{Langer12} and \citet{Smith14}. 

\section{SINGLE STAR MODELS}

\subsection{Mass Loss and Final Mass}\label{sect:singlemass}

It has been widely believed that Wolf-Rayet (WR) stars  are observational
counterparts of SNe Ib/c progenitors \citep[e.g.][]{Meynet03, Massey03,
Crowther07, Smartt09}.  Although helium stars as WR stars can be produced by
binary interactions~\citep[e.g.,][]{Petrovic05a, Vanbeveren07}, a large
fraction of WR stars are found in isolation \citep{Hucht01, Crowther07}, and
must have been produced from massive single stars\footnote{Some massive single
stars on the main sequence may be products of binary mergers, but here we do
not distinguish them from singly-formed massive stars}  via mass loss due to
stellar winds (The so-called Conti scenario; \citealt{Conti76}).  

Evolutionary models of massive stars with mass loss predict that there exists
an initial mass limit for WR stars, above which stars can lose the entire
hydrogen envelope during the post main sequence phases \citep[e.g.][]{Maeder87,
Schaller92, Vanbeveren98, Meynet03, Eldridge06, Georgy12}.  A useful constraint
on this mass limit can be provided by galactic WR stars.  Observations indicate
that WR stars of WN type in our Galaxy have the lower bolometric luminosity
limit of $\log L/\mathrm{L_\odot} \simeq 5.3$~\citep{Hamann06}.  This roughly
corresponds to 10~\Msun{} of a naked helium star, which requires an initial
mass of about 25~\Msun{}.  Stellar evolution models indicate that non-rotating
stars at solar metallicity cannot lose their hydrogen envelope to become a WR
star if $M_\mathrm{ZAMS} < 40$~\Msun{}, with the most commonly adopted mass
loss rate from red supergiant stars given by \citet{deJager88}.  Enhancement of
mass loss due to rotation or pulsation compared to the de Jager rate and some
alternative empirical mass loss prescriptions have been invoked to resolve this
discrepancy \citep[e.g.][]{Vanbeveren98, Salasnich99, Meynet03, Loon05, Loon08,
Yoon10a, Ekstroem12}. 

\begin{figure}
\begin{center}
\includegraphics[width=\columnwidth]{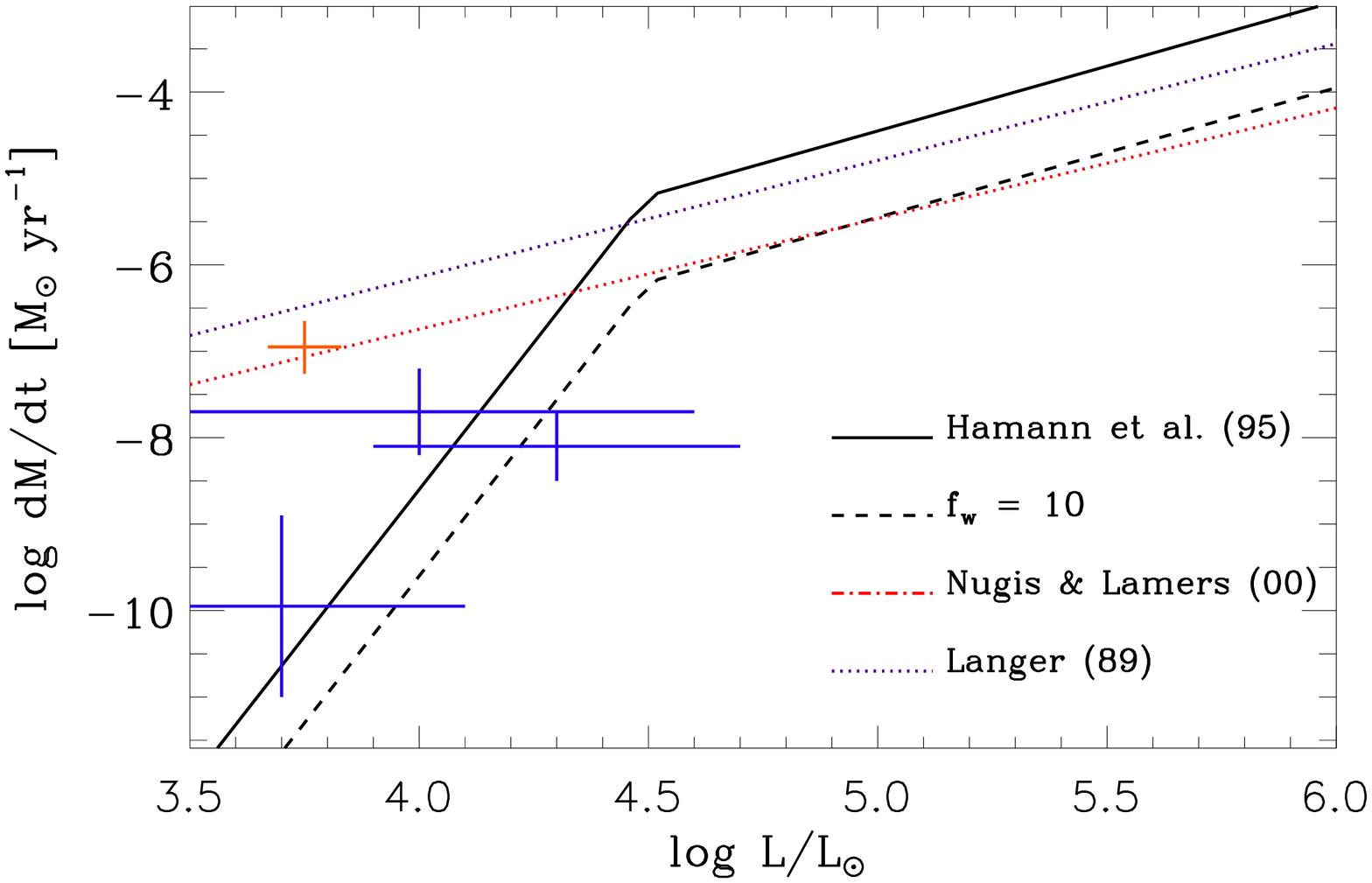}
\caption{Comparison of different mass loss prescriptions of massive helium stars on the zero-age helium main sequence 
as a function of the surface luminosity, which are based on WR stars ($\log L/\mathrm{L_\odot} > 4.5 $). 
The dot-dashed line and the dotted line give the WR mass loss rates by \citet{Nugis00} and \citet{Langer89b}, respectively. 
The solid line denotes the mass loss rate prescription given by Eq.~(1): 
 the WR mass loss rate  by \citet{Hamann95} for $\log L/\mathrm{L_\odot} \ge 4.5 $ and 
the mass loss rate of relatively low-mass helium stars for $\log L/\mathrm{L_\odot} < 4.5 $, 
which is based on the extreme helium stars analyzed by \citet{Hamann82}. 
The blue data points with the error bars are the mass loss rates of these extreme helium stars. 
The  orange point with the error bars denotes the mass loss rate of the quasi-WR star HD 45166 \citep{vanBlerkom78, Groh08}.  
The dashed line is 10 times lower than the solid line:  $f_\mathrm{w}$ is the reduction factor compared
to the mass loss rate given by Eq.~(1). 
}\label{fig:wrwind}
\end{center}
\end{figure}

Once a star becomes a WR star, further mass loss due to WR winds determines its
final mass. In the 80s and 90s, a fixed value of about $3 -
8\times10^{-5}~\mathrm{M_\odot~yr^{-1}}$ or mass-dependent values have been
widely used for the WR mass loss rate in most evolutionary models
\citep{Maeder87, Langer89b, Schaller92, Schaerer93a, Schaerer93b, Meynet94,
Woosley93, Woosley95}.  Later studies began to consider the WR mass loss rate
as a function of the luminosity and the surface abundances of helium and metals
in a more explicit way \citep[e.g.,][]{Wellstein99, Meynet03,
Meynet05, Eldridge06,  Georgy12}.  More important, with the growing evidence
for hydrodynamic clumping of WR wind material, recent estimates for the WR mass
loss rate give significantly lower values than previously thought  \citep[e.g.,][see
Fig.~1][]{Nugis00, Hamann06, Crowther07, Sander12}. Several different prescriptions for the WR
mass loss rate are compared in Fig.~\ref{fig:wrwind}.  

\begin{figure}
\begin{center}
\includegraphics[width=\columnwidth]{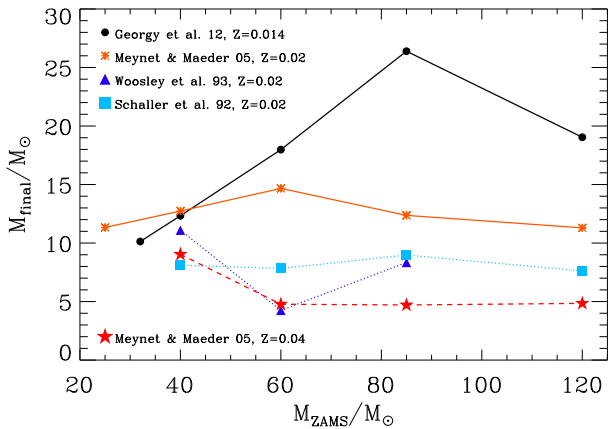}
\caption{
Theoretical predictions on the final mass of single star progenitors for SNe Ib/c, 
as a function of the initial mass (i.e., mass on the zero-age main sequence). 
Circle: rotating models of \citet{Georgy12} at $Z = 0.014$, 
Asterisk: rotating models of \citet{Meynet03} at $Z = 0.02$, 
Star: rotating models of \citet{Meynet05} at $Z = 0.04$ with a metallicity dependent WR mass loss rate.
Triangle: non-rotating models of \citet{Woosley93} at $Z = 0.02$, 
Square: non-rotating models of \citet{Schaller92} at $Z = 0.02$. 
}\label{fig:finalmass}
\end{center}
\end{figure}

The single star models produced later than 2000 predict
systematically higher final masses of SN Ib/c progenitors  than those in the
80s and 90s, as summarized in Fig.~\ref{fig:finalmass}. For example, with the
Langer's mass-dependent WR mass loss rate~\citep{Langer89b}, a 60~\Msun{} star
at solar metallicity can become a SN Ib/c progenitor with an final mass as low
as 4.25~\Msun{}~\citep{Woosley93}. By contrast the models with the WR mass loss
rate of \citet{Nugis00} give final masses higher than 10~\Msun{}, at solar
metallicity. 

This high final mass ($M_\mathrm{f} > 10$~\Msun) has consequences on the SN
explosion.  First of all, such massive helium stars have large amounts of
binding energy.  This would make successful explosion of these progenitors
difficult: they  may collapse to a black hole, without making an ordinary SN
Ib/c \citep[e.g.][and references therein]{Heger03}.  Secondly, even if they
exploded successfully, the resultant light curves would be too broad to be
compatible with observations \citep{Woosley93, Woosley95, Dessart11, Drout11,
Cano13, Lyman14, Taddia14}. This brings into question the importance of single
WR stars as progenitors of SNe Ib/c at solar metallicity.  However,  the role
of single stars at super-solar metallicity may be significant given their
relatively low final masses (Fig.~\ref{fig:finalmass}).

\subsection{Surface Properties}\label{sect:singlesurface}

\begin{figure}
\begin{center}
\includegraphics[width=\columnwidth]{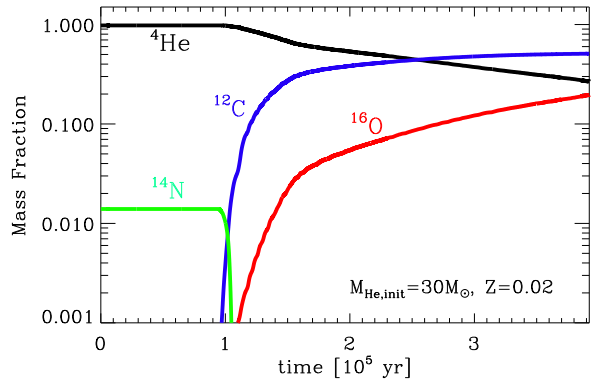}
\caption{Evolution of the chemical composition at the surface 
of a 30~\Msun{} helium star at $Z = 0.02$, with the WR mass loss 
rate by \citet{Nugis00}. The calculation was terminated at the end of core neon burning. }\label{fig:surface}
\end{center}
\end{figure}

WR stars have large convective cores, being close to the Eddington limit, and
rapidly lose the outer helium-rich layers by WR winds.    This makes them
almost chemically homogeneous.   Therefore, stellar evolution models predict
that WR stars evolve systematically bluewards on the HR
diagram~\citep[e.g.,][]{Georgy12, Yoon12, Eldridge13},  in contrast to the case
of the evolution of hydrogen-rich stars which evolve redwards in general. The
surface composition of chemical elements also  evolves. First, as the residual
of the hydrogen envelope is removed by winds, they evolve from WNL type to WN
type. As they lose more mass, the products of helium burning including carbon
and oxygen begin to appear at the surface, to become WC and WO stars
(Fig.~\ref{fig:surface}). The general consensus is that  WR stars from
sufficiently high initial masses  evolve according to the following order:  WNL
$\rightarrow$ WN $\rightarrow$ WC $\rightarrow$ WO. 

\begin{figure}
\begin{center}
\includegraphics[width=\columnwidth]{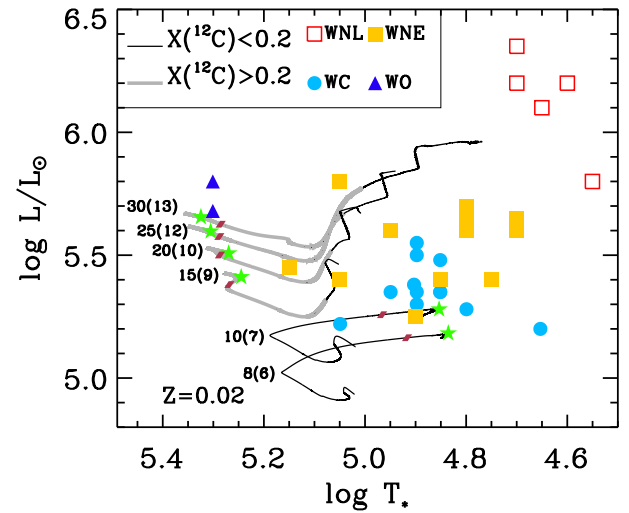}
\caption{Evolution of massive helium stars at solar metallicity compared to the observed Wolf-Rayet stars in our galaxy on the Hertzsprung-Russel diagram.
The WR mass loss rate prescription by \citet{Nugis00} was adopted in the evolutionary models. The initial mass for each evolutionary track
is marked by the label in the left hand side, and the final mass is indicated in the parenthesis. The thick grey lines mark the evolutionary stage 
where the surface mass-fraction of carbon is higher than 0.2. 
The star symbol denotes the end point of the evolution, which is the end of core neon burning. 
This figure is a reproduction of Figure~3 in \citet{Yoon12} with 
permission from Astronomy \& Astrophysics, \copyright~ESO. 
}\label{fig:hr}
\end{center}
\end{figure}

In this scenario, WR stars should become more compact as they evolve from WNL
to WO.   Indeed, WNL and WO stars in our galaxy have lowest and highest surface
temperatures, respectively, in agreement with the theoretical prediction
\citep{Hamann06, Sander12}. But stellar evolution models have great difficulty
in explaining many of the surface properties of WR stars.  In particular, the
observed WR stars are found to have much larger radii and lower surface
temperatures than what the evolutionary models predict
\citep[e.g.,][]{Hamann06, Crowther07, Sander12}.  The reason for this
discrepancy is not well understood yet. 
Inflation of the envelope with a density inversion is observed in
WR star models near the Eddington limit in hydrostatic equilibrium~\citep{Ishii99, Petrovic06}, 
but this is still not sufficient to fit observations as seen in Fig.~\ref{fig:hr}. 
A recent suggestion is that the
observationally implied inflation of WR stars may result from density
inhomogeneities and the consequent enhancement of opacity in the sub-surface
convective layer  \citep{Graefener12}.  

This envelope inflation affects the bolometric correction, making WR stars
fairly luminous in the optical ($M_\mathrm{V} \lesssim -4$). If the optical
luminosities of the observed WR stars represented those of SN Ib/c progenitors at
the pre-SN stage, the previous search for a SN Ib/c progenitor in pre-SN images
would have been successful~\citep{Maund05a, Maund05b, Crockett07, Smartt09,
Eldridge13}.  As of today, only one tentative identification  has been reported
with  the SN Ib iPTF13bvn~\citep{Cao13}.  All of the other searches for SN Ib/c
progenitors have failed, even for the case with a very deep detection limit
($M_\mathrm{V} \gtrsim -4.3$; \citealt{Crockett07, Eldridge13}).  This result
has often been interpreted as evidence for binary star progenitors
\citep{Crockett07, Smartt09}.  

It should be noted that the majority of
the observed WR stars must be on the helium main sequence, which is still far
from the final evolutionary stage.  After core helium exhaustion, the evolution
of the core in a WR star is dominated by neutrino cooling and undergoes rapid
Kelvin-Helmoltz contraction.  With a sufficient amount of helium in the
envelope, this would lead to further expansion of the helium envelope due to
the so-called mirror effect. However, single WR stars would rapidly lose helium
in the envelope as implied by the high mass loss rate, and the overall radius
would gradually decrease as the stellar evolution models predict. At the pre-SN
stage, many WR stars would tend to become very hot, and optically faint like WO
stars despite their very high bolometric luminosities \citep{Yoon12}.
This means that the non-detection of most SN Ib/c progenitors in the previous
attempts does not necessarily exclude single WR progenitors, and other
constraints like  ejecta masses of SNe Ib/c  should also be taken into account
to better understand the nature of SNe Ib/c progenitors.

\subsection{Helium}\label{sect:singlehe}

 The production of He~I lines is found to depend both on the total He
mass~\citep{Hachinger12} and on the helium distribution in the
envelope~\citep{Dessart11, Dessart12}.  Non-thermal excitation and
ionization of helium also play the key role for the formation of helium lines in
SNe Ib~\citep{Lucy91, Woosley97, Dessart12, Hachinger12}.  This does not only
require presence of helium in the progenitors, but also  strong chemical mixing
between helium in the envelope and radioactive $^{56}$Ni produced in the
innermost region of the SN ejecta~\citep{Dessart12, Hachinger12}.  We still do
not know exactly how much helium is needed for SNe Ib.   This limit must depend
on the degree of mixing of helium and nickel, which may in turn depend on the
detailed structure of the progenitor and the energy and asymmetry of the
explosion. Many authors simply assume a certain amount of helium (e.g., 0.5 -
0.6~\Msun{}) as the lower limit for SN Ib progenitors
\citep[e.g.][]{Wellstein99, Yoon10, Georgy12}.  Recently \citet{Hachinger12}
suggested 0.14~\Msun{} as the maximum possible amount of helium that can be
hidden in the SN spectra, based on a spectroscopic study of several SN Ib/c
with relatively low inferred ejecta masses. 

\begin{figure}
\begin{center}
\includegraphics[width=\columnwidth]{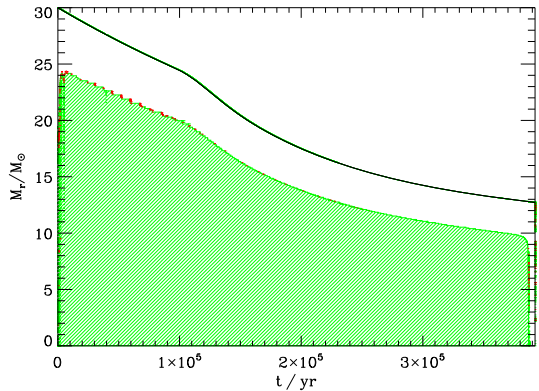}
\caption{Evolution of the internal structure
of a 30~\Msun{} helium star, for which 
the Nugis \& Lamers' WR mass loss rate was adopted. The helium-burning
convective core is marked by the hatched lines. 
The black solid line marks 
the surface of the star. The calculation was
terminated at the end of core neon burning.}\label{fig:kipp1}
\end{center}
\end{figure}

In the most recent single star models \citep{Georgy12}, the total amounts of
helium in SN Ib/c progenitors range from 0.28~\Msun{} to 2.2~\Msun{}. This is
significantly higher than the proposed limit of 0.14~\Msun{} by Hachinger et
al.  In fact, helium mass as low as 0.14~\Msun{} is very difficult to achieve
with stellar evolution models.  \citet{Woosley93} found that a 60~\Msun{} star
can become a 4.25~\Msun{} SN Ib/c progenitor with a WR mass loss rate much
higher than nowadays adopted, but even in this extreme case, the remaining
helium mass was as large as 0.18~\Msun{}.  The reason for this difficulty is
largely related to the dynamical adjustment of the stellar structure of WR
stars with mass loss. As shown in Fig.~\ref{fig:kipp1} as an example, the size
of the helium-burning convective core in a WR star decreases as the WR star
loses mass by winds (Fig.~\ref{fig:kipp1}), and therefore some amount of helium
can remain unburned until the end of core helium burning even if more than half of
the initial mass is lost.  The residual helium could be completely removed with
efficient mass loss during the later evolutionary stages.  The current models
predict, however, that the effect of mass loss during the post helium burning
phases is relatively minor mainly because of the relatively short evolutionary
time.  We discuss the problem of helium in SNe Ib/c progenitors in
Sect.~\ref{sect:binaryhe} in more detail.

The mass fraction of helium in the outermost layers can also play an important
role for the early time light curves and spectra of SNe Ib/c.
\citet{Dessart11} showed that if the helium mass fraction is sufficiently large
($\sim 0.9$),  He I lines can be produced without the contribution of
non-thermal processes for several days after the shock breakout, while no
helium lines are seen for a low helium mass fraction ($\lesssim 0.5$) even with
a total helium mass of about 1~\Msun{}.  
Therefore, He I lines during the early epoch
of a SN Ib/c  will provide an important constraint on the progenitor.  
Recent single star models in the
literature predict that helium mass fraction at the surface of SN Ib/c
progenitors is below 0.4 except for a limited initial mass range above the
critical mass for WR star progenitors \citep[e.g.,][]{Meynet03, Meynet05,
Georgy12}, in contrast to the case of binary star models that predict the
majority of SN Ib/c progenitors have a surface helium mass fraction higher than
0.9 (Sect.~\ref{sect:binaryhe}).  
Given that the total amount of helium is also systematically smaller 
in single star models than in binary star models as discussed below, 
this implies that single star evolution is probably prone to SNe Ic.

\subsection{Rotation}\label{sect:singlerot}

 Super-luminous supernovae (SLSNe) and very energetic explosions like
gamma-ray bursts (GRBs) can be driven by rapid rotation.  The most commonly
invoked mechanisms for these events include the collapsar
scenario~\citep{Woosley93, MacFadyen99} and the magnetar
scenario~\citep[e.g.][]{Wheeler00, Burrows07, Kasen10, Woosley10}.  So far, all
of the supernovae associated with GRBs belong to Type Ic~\citep{Woosley06,
Hjorth13}, and many SLSNe are also found to be SNe Ic.  While the collapsar
scenario  still remains most popular to explain GRBs~\citep{Woosley06b}, the
magnetar-driven explosion is nowadays the most invoked mechanism  for the SLSNe
Ic \citep[e.g.,][]{Chomiuk11, Dessart12b,  Inserra13, Nicholl13, Mazzali14}.
The contending mechanism for SLSNe Ic is the pair-production
instability~\citep[e.g.][]{Barkat67, GalYam09, Kozyreva14}.

It is still a matter of debate which evolutionary channels of SNe Ib/c
progenitors can lead to rotation-driven explosions like GRBs and SLSNe-Ic.
Observations indicate that the majority of massive stars on the main
sequence are rapid rotators, where the necessary condition for both collapsar
and magnetar mechanism could be fulfilled if they retained the angular momentum
until the pre-collapse stage~\citep[e.g.,][]{Heger00}.  However, massive stars
may undergo angular momentum redistribution via mass loss due to stellar winds
and/or binary interactions, and the transport of angular momentum
\citep{Maeder00, Heger00, Hirschi04, Heger05, Petrovic05b, Yoon10} . The
angular momentum transfer may occur on a dynamical timescale in convective
layers by convection.  In radiative layers, rotationally-induced hydrodynamic
instabilities like the shear instability and Eddington-Sweet circulations may
transport angular momentum \citep{Maeder00}.  Dynamo actions may also occur in
radiative layers according to the so-called Tayler-Spruit dynamo theory
\citep{Spruit02}, which may cause strong magnetic torques across differentially
rotating layers.

Theoretical studies indicate that, without magnetic fields, angular momentum
transport is severely inhibited by the chemical stratification across the
boundary between the stellar core and the hydrogen envelope ($\mu$-barrier;
\citealt{Meynet97, Heger00}).  Single star progenitors of SNe Ib/c can thus
retain a significant amount of angular momentum until the pre-SN stage, even
though most of the initial angular momentum is lost by stellar winds
\citep{Heger00, Hirschi04}: the predicted amounts of angular momentum in the
cores are much more than what neutron stars can have at the break-up velocity,
and enough to produce a long gamma-ray burst (GRB) within the collapsar
scenario \citep{Woosley93a}.   This means that almost all SN Ib/c progenitors
have enough angular momentum to form  GRB/SLSN-Ic progenitors. 
Given that rapid rotation is only one necessary condition for collapsar/magnetar
production, this should not necessarily lead to the
conclusion that non-magnetic models predict too many  GRBs and SLSNe-Ic
compared to the observation.  However, because such exotic explosions belong to
a subset of SNe Ib/c, the result of non-magnetic models implies that GRBs and SLSNe-Ic should
occur more frequently at higher metallicity as ordinary SNe Ib/c do, if mass
loss by winds provided the main evolutionary path for their progenitors.
Contrary to this expectation, observations indicate that low-metallicity is
preferred for both GRBs and SLSNe-Ic \citep[e.g.,][]{Modjaz08, Graham13,
Lunnan14}.

Magnetic torques can easily overcome the hindrance by the chemical
stratification to the transport of angular momentum.  Magnetic models with the
Tayler-Spruit dynamo predict that  single WR stars rotate too slowly to produce
a magnetar/collapsar \citep{Heger05}.  This is consistent with
the fact that GRBs and SLSNe-Ic are very rare.  Magnetic models also better
explain the spin rates of young millisecond pulsars. 

Several authors have questioned the validity of the Talyer-Spruit dynamo
theory~\citep{Zahn07, Gellert08} and we still cannot draw any solid  conclusion
on which case between magnetic and non-magnetic models better represents the
reality.  However, it is not only massive stars but also intermediate- and
low-mass stars that provide evidence for very efficient transport of angular
momentum in the radiative layers.  Such examples include slow rotation of
isolated white dwarfs~\citep{Suijs08}, the radial velocity profile in the
Sun~\citep{Eggenberger05}, and recent asteroseismic results of low-mass
stars~\citep{Eggenberger12, Cantiello14}.  Our tentative conclusion is that
some strong braking mechanism like the Tayler-Spruit dynamo is actually working
in stars, and that magnetic models may better explain recent observations in
general.  The role of magnetic fields on the evolution of stars still
remains a very challenging subject of future study. Some other mechanisms like
baroclinic instability~\citep{Fujimoto93} and pulsational
instabilities~\citep{Townsend08} may also play an important role for the
transport of angular momentum, but  have not been extensively studied for
massive stars yet.

\section{BINARY STAR MODELS}\label{sect:binary}

\begin{sidewaysfigure*}
\centering
\scalebox{0.85}
{\includegraphics[angle=90]{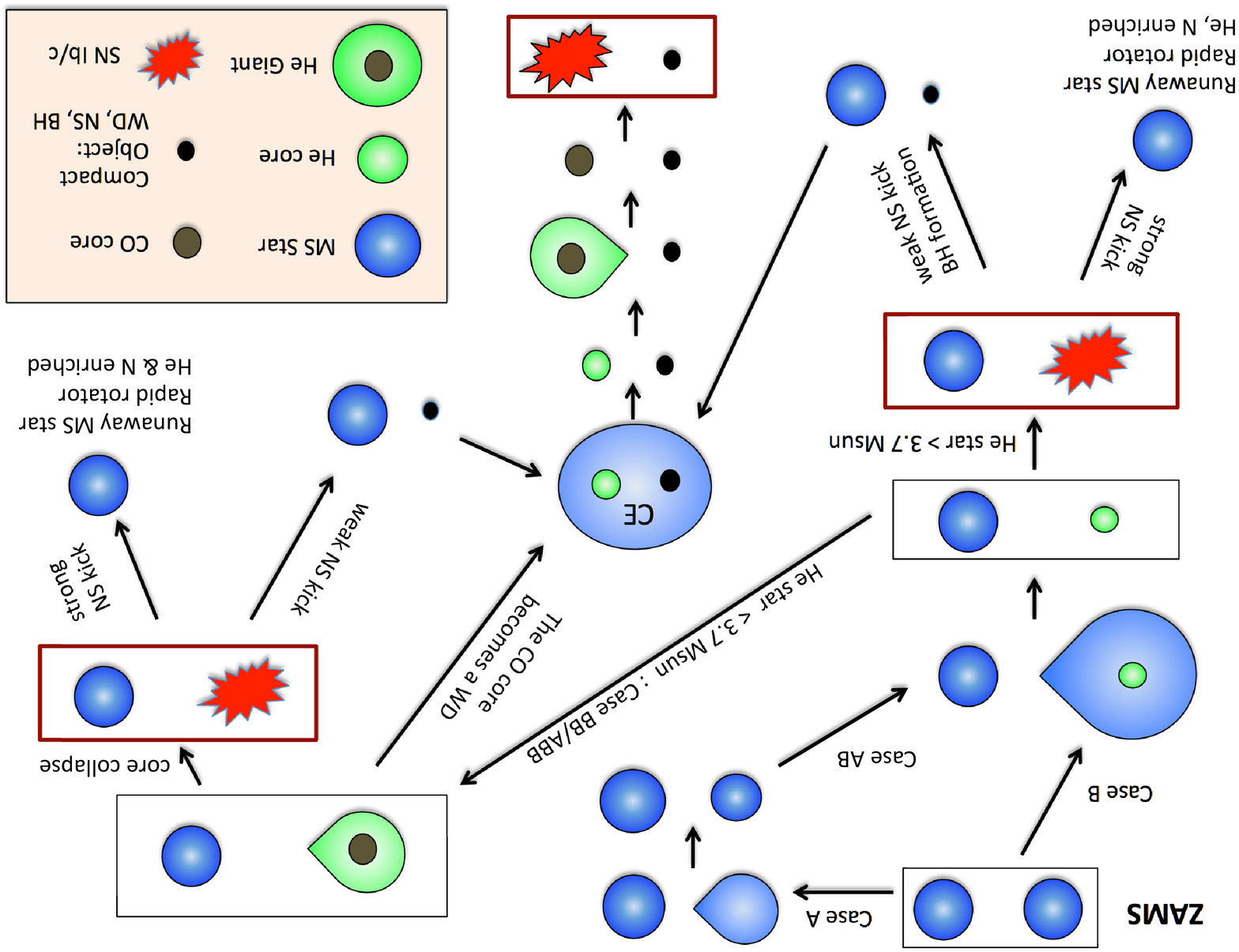}}
\caption{Evolutionary paths of a massive binary system towards a type Ib/c supernova.
The bifurcation points and the end points of the evolution are marked by square boxes. 
}\label{fig:binary}
\end{sidewaysfigure*}

\subsection{Binary evolution towards a SN Ib/c}\label{sect:binaryevol}

The majority of massive stars form in binary systems~\citep[e.g., see][for a
recent observational analysis on the population of binary systems]{Sana12}. A
large fraction of them are believed to experience binary interactions during
the course of their evolution, mainly due to the increase of stellar radius.
Once the more massive star (the primary star) fills the Roche lobe in a binary
system, mass transfer to the less massive star (the secondary star) begins.
Mass transfer can be unstable if the mass ratio of the stellar components
(i.e., $q = M_2/M_1$ where $M_1$ and $M_2$ are the masses of the primary and
secondary stars, respectively) is sufficiently small. Unstable mass transfer
will lead the binary system to a contact phase, which may eventually make the
stellar components merge to become a single star.  Although binary mergers are
related to many important topics like stellar rotation, peculiar SNe and long
gamma-ray bursts~\citep{Fryer05, deMink14, Justham14}, here we focus our
discussion on non-merging systems that can produce ordinary SNe Ib/c. Recent
analyses also indicate that the fraction of stable binary systems is much
higher than previously believed~\citep{Kobulnicky07, Sana12}, and thus the
event rate of SNe Ib/c can be well explained by binary progenitors
\citep{Kobulnicky07, Smith11}.

In the literature, mass transfer is often categorized into Case A, Case B and Case C, 
depending on the evolutionary stage of the primary star when it fills the Roche lobe~\citep{Kippenhahn67, Lauterborn70}, 
as the following: 
\begin{itemize}
\item Case A mass transfer : mass transfer during the main sequence. 
\item Case B mass transfer : mass transfer during the helium core contraction phase.
\item Case C mass transfer : mass transfer during the core helium burning and later evolutionary stages.
\end{itemize}
Some binary systems may undergo multiple mass transfer phases, depending on the
initial orbital periods and masses of the stellar components. The mass transfer phases
that follow Case A/B mass transfer are often denoted as the following: 
\begin{itemize}
\item Case AB mass transfer :  mass transfer from the primary star
that has previously undergone the Case A mass transfer, during the helium core contraction phase.  
\item Case BB mass transfer : mass transfer from the primary star that has previously undergone Case B mass transfer,  
during the late evolutionary stages (mostly  after core helium exhaustion for SN Ib/c progenitors). 
\item Case ABB mass transfer :  mass transfer from the primary star that 
has previously undergone Case AB mass transfer, during the late evolutionary stages (mostly  after core helium exhaustion for SN Ib/c progenitors). 
\end{itemize}

\begin{figure}
\begin{center}
\includegraphics[width=\columnwidth]{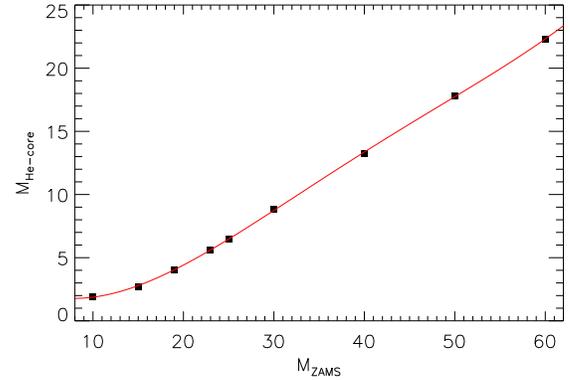}
\caption{The helium core mass at the terminal age of the main sequence as a function of the initial mass for 
single stars. Based on non-rotating models without overshooting.}\label{fig:mimf}
\end{center}
\end{figure}

Helium stars as SN Ib/c progenitors can be made via Case B/AB mass
transfer as illustrated in Fig.~\ref{fig:binary}.  The initial mass of such a
helium star corresponds to the helium core mass ($M_\mathrm{He-core}$) of the
primary star at the onset of Case B/AB mass transfer.  For Case B systems,
$M_\mathrm{He-core}$ can be given by a well-defined function of the ZAMS mass of
the primary star as shown in Fig.~\ref{fig:mimf}. 

The evolution of the helium stars produced via Case B/AB mass transfer largely
depends on their masses.  Stellar evolution models indicate that Case BB/ABB
mass transfer occurs when the primary star becomes a helium giant if the helium
star mass is initially less than about 3.5 - 4.0~\Msun{}, depending
on the adopted mass loss rate and metallicity.  More massive helium stars do not
interact anymore with the secondary stars after Case B/AB mass transfer, but
can still lose mass further by winds. WR winds may be induced if the helium
star mass is sufficiently high ($\gtrsim 10$~\Msun) but the mass loss rate from
less massive helium stars is not well constrained observationally because such
relatively low-mass helium stars have been rarely observed
(see Sects.~\ref{sect:binarymass} and~\ref{sect:counterpart} below for more discussions).  

It is not only the primary star, but also the secondary star that can produce
a SN Ib/c. In a close binary system, the primary star will leave a compact star
remnant if it explodes as a SN Ib/c via Case B/AB/BB/ABB mass transfer, or if
it becomes a white dwarf via Case BB/ABB mass transfer (see
Fig.~\ref{fig:binary}).  Unless the binary system is unbound by a strong
neutron star kick, it will form a common envelope after the core hydrogen
exhaustion in the secondary star.   A short-period binary system consisting of a
helium star plus a compact star (white dwarf, neutron star, or black hole) will
be produced after the common envelope ejection. Explosion of the helium star
will then produce a SN Ib/c.  

Note also that Fig.~\ref{fig:binary} still does
not depict all the possible binary paths for SNe Ib/c, and there may exist
other relatively rare channels.  For example, some authors found that with Case
A mass transfer, the SN order can be reversed for a limited parameter
space: the secondary star first explode as a SN Ib/c, followed by SN Ib/c
explosion of the primary star as a helium star in isolation or in a compact
binary system with a neutron star companion, depending on the impact of the
neutron star kick  \citep{Pols94, Wellstein01}. 

The contribution of each evolutionary path to the total production of SNe Ib/c
may depend on several physical parameters.  They include the so-called mass
accretion efficiency (i.e,  the ratio of the accreted mass onto the secondary
star to the transferred mass from the primary star), the specific angular
momentum of any matter that is not accreted on the secondary star but lost from
the binary system, distribution of neutron star kick velocities, and the common
envelope ejection efficiency \citep[e.g.][]{Podsiadlowski92}. It is beyond the
scope of this review to discuss all the details regarding binary population
with respect to SNe Ib/c, and readers are referred to \citet{Podsiadlowski92},
\citet{Izzard04}, \citet{Eldridge08, Eldridge11, Eldridge13}, as well as  the
contribution by S.E. de Mink in this issue.  Here, it may be sufficient to say
that  the dominant channel to SNe Ib/c in binary systems is the Case B/BB,
among others~\citep{Podsiadlowski92}.  In the sections below, we focus on the
detailed properties of  SN Ib/c progenitors predicted from evolutionary models.

\subsection{An Example of Binary Models}\label{sect:binaryexample}

\begin{figure*}[ht]
\begin{center}
\includegraphics[width=0.8\textwidth]{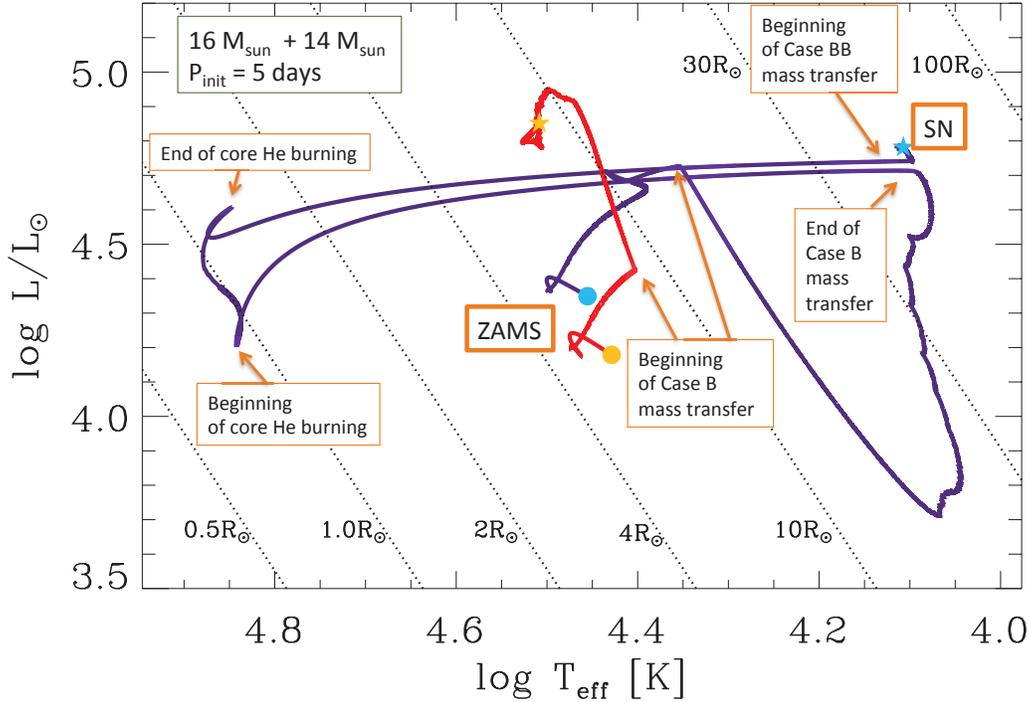}
\caption{Evolution of a binary system  consisting of 16~\Msun{} plus 14~\Msun{} stars with the initial period of 5 days.
on the Hertzsprung-Russel diagram. The evolutionary tracks of the primary and secondary stars are marked
by dark-blue and red colors, respectively. The adopted mass loss rate for helium stars 
is given by Eq.~(1), reduced by a factor of 5 ($f_\mathrm{w} =5$). The initial and final points
of each track are marked by the filled circle and the star symbol, respectively}\label{fig:hr2}
\end{center}
\end{figure*}

To model the evolution of a binary system, we have to consider the change of
the orbit due to stellar winds, mass transfer and/or gravitational wave
radiation, and mass exchange between the stellar components via mass transfer.
To investigate the effect of rotation, angular momentum exchange between stars
and the orbit via tidal synchronization and the spin-up effect of the secondary
during the mass transfer phases should also be followed ~\cite[see][for a
recent review]{Langer12}.  

Many evolutionary models of massive binary stars
have been presented in the literature, but only a limited number of studies aimed
at detailed investigation of the structure of SNe Ib/c progenitors  near/at the
pre-SN stage~\citep{Woosley95, Wellstein99, Yoon10, Eldridge13} .
Before we summarize the main results of these studies, we give an example for
the evolution of relatively low-mass SN Ib/c progenitors, which is very
different from that of massive WR stars. 

\begin{figure}
\begin{center}
\includegraphics[width=\columnwidth]{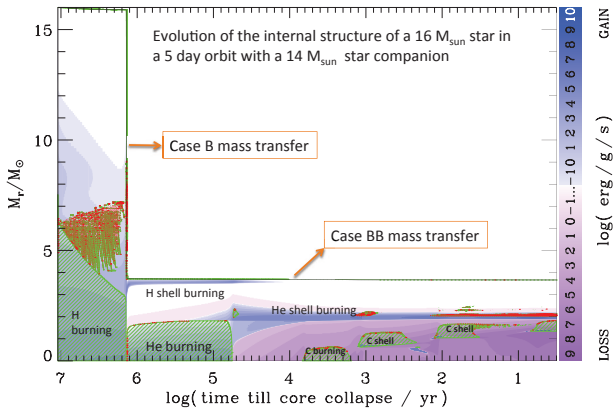}
\caption{Evolution of the internal structure of the primary star in 
a binary system consisting of 16~\Msun{} plus 14~\Msun{} stars with the initial period of 5 days.
Convective layers are marked by green hatched lines, and semi-convection layers by red dots. 
The surface of the star is indicated by the black solid line. 
The blue and pink color shading marks net energy gain or loss 
from nuclear energy generation and neutrino emission.}\label{fig:kippen}
\end{center}
\end{figure}

The evolution of a SN Ib/c progenitor having $M_\mathrm{ZAMS} = 16$~\Msun{} in
a Case BB system is illustrated in~Figs.~\ref{fig:hr2} and~\ref{fig:kippen}.
After Case B mass transfer, the primary star becomes a hot and compact helium
star. The surface hydrogen and helium mass fraction at this stage is about 0.28
and 0.7, respectively and a small amount of hydrogen of about 0.05~\Msun{} is
still retained in the outermost layer. As a result, the convective helium
burning core can  grow with hydrogen shell burning even though the total mass
somewhat decreases due to mass loss by winds (Fig.~\ref{fig:kippen}). This is
contrasted to the case of mass-losing pure helium stars where the convective
helium burning core shrinks in size in terms of the mass coordinate
(Fig.~\ref{fig:kipp1}).  During the post-helium burning stages, the helium
envelope rapidly expands and Case BB mass transfer is initiated when carbon
burning begins in the core (Figs.~\ref{fig:hr2} and~\ref{fig:kippen}).  The SN
explosion will occur when the surface temperature becomes fairly low ($\log
T_\mathrm{eff} \simeq 4.1$, Fig.~\ref{fig:hr2}) while the star is still
undergoing Case BB mass transfer.  This is in stark contrast to the case of
massive WR star progenitors that evolve bluewards throughout and explode when
they become very hot ($\log T_\mathrm{eff}  > 5~\mathrm{K}$;
Fig.~\ref{fig:hr}). The helium envelope expansion makes this binary progenitor
bright in the optical bands compared to the case of WR progenitors, as
discussed below (Sect.~\ref{sect:binarysurface}).  Of course, more massive
progenitors having $M_\mathrm{ZAMS} \gtrsim 30$ will become a WR star even in
binary systems after Case B/AB transfer, and evolve like a single WR star
thereafter.

\subsection{Mass Loss and Final Mass}\label{sect:binarymass}

As mentioned above, one of the biggest uncertainties in the evolution of SN
Ib/c progenitors is the mass loss rate of naked helium stars. The mass loss
rate of WR stars ($\log L/\mathrm{L_\odot} \gtrsim 5$)  is relatively well
known, but less luminous helium stars have not been well studied
observationally.  Several authors therefore used extrapolated values of the
Nugis \& Lamers rate or the Langer rate (see Fig.~\ref{fig:wrwind}) for the whole possible range of helium
star mass \citep{Woosley95, Eldridge08, Eldridge13}.  
On the other hand, \citet{Braun97} and \citet{Wellstein99} used a
significantly reduced mass loss rate for $\log L/\mathrm{L_\odot} < 4.5$ as the
following:  
\begin{eqnarray}
\log \left(\frac{\dot{M}}{\mathrm{M_\odot~yr^{-1}}}\right)  = & \nonumber   \\ 
   \begin{cases}
     - 11.95 + 1.5 \log L/\mathrm{L_\odot}  &  \text{for}~ \log L/\mathrm{L_\odot} \ge 4.5 \\
     - 35.8 + 6.8 \log L/\mathrm{L_\odot}   &  \text{for}~ \log L/\mathrm{L_\odot}  <  4.5 ~. 
  \end{cases}& 
\end{eqnarray}\label{eqhewind}
Here the WR mass loss rate for $\log L/\mathrm{L_\odot} \ge 4.5$ is given by \citet{Hamann95}. 
The prescription for less luminous helium stars ($\log L/\mathrm{L_\odot} < 4.5$)
is based on the observations of extreme helium stars by \citet{Hamann82}. 
Fig.~\ref{fig:wrwind} indicates that the simple extrapolation of the WR mass loss rate  
down to  $\log L/\mathrm{L_\odot} <  4.5$
may lead to a significant overestimate
even with the Nugis \& Lamers rate, which is about 10 times lower than 
that of \citet{Hamann95} for  $\log L/\mathrm{L_\odot} \ge 4.5$. 

Given that  the stellar wind mass loss rates of massive stars were usually
overestimated before the late-90s (see Sect.~\ref{sect:singlemass} above),
mass-loss rates reduced by a certain factor ($f_\mathrm{w}$) compared to that
of Eq.~(1) were applied for some binary models of \citet[][$f_\mathrm{w} =
2$]{Wellstein99}  and for all the models of \citet[][$f_\mathrm{w} =
5~\mathrm{or}~10$]{Yoon10}.  Note that the case of $f_\mathrm{w} = 10$, for
which the mass loss rate becomes comparable to the Nugis \& Lamers rate with
$\log L/\mathrm{L_\odot} \ge 4.5$ , is still compatible with the observation of
extreme helium stars of $\log L/\mathrm{L_\odot} < 4.5$ (Fig.~\ref{fig:wrwind}).  The
caveat is that the extreme helium stars in the figure
are giant stars at a pre-white dwarf stage, having masses of only about
0.8~\Msun{}~\citep{Jeffery10}, and cannot represent ordinary SN Ib/c progenitors.  

\begin{figure}
\begin{center}
\includegraphics[width=\columnwidth]{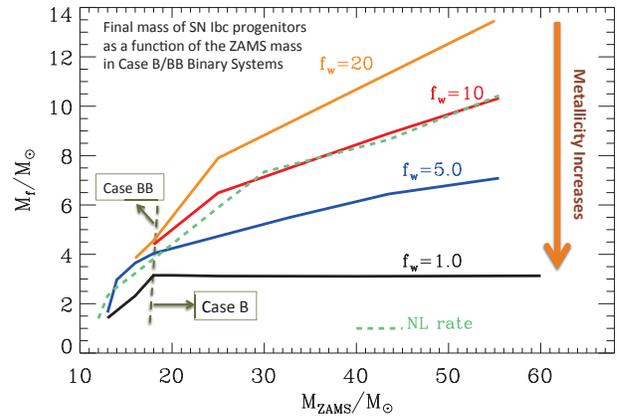}
\caption{ The final mass of SN Ib/c progenitors via Case B/BB mass transfer as a function of the initial mass, 
for different helium star mass loss rates.  Here  $f_\mathrm{w}$ denotes the reduction factor that are applied 
to the mass loss rate given by Eq.~(1): $f_\mathrm{w} =$ 10 and 20 roughly corresponds to solar and SMC metallicity, respectively. 
The presented results are based on  the full binary models by \citet{Wellstein99} for $f_\mathrm{w} = 1$ and 
the binary models and pure helium star models by \citet{Yoon10} for the others. The boundary 
line between Case B and BB systems is marked by the dashed line.
The data with $M_\mathrm{ZAMS} \le 25$~\Msun{} in \citet{Yoon10} is based on the full binary evolution calculations
but it is based on pure helium star models for  $M_\mathrm{ZAMS} > 25$~\Msun{} assuming 
that the pure helium star was produced via Case B mass transfer from the primary star with the corresponding ZAMS mass in a binary system. 
See \citet{Yoon10} for more details.    
The result with the WR mass loss rate by \citet{Nugis00} at solar metallicity 
from unpublished binary star models (Yoon, S.-C., in prep) is marked by the green dashed line, 
for which the boundary for Case B and BB shifts to about $M_\mathrm{ZAMS} =  15$~\Msun{}.  
}\label{fig:finalmass2}
\end{center}
\end{figure}

On the other hand, the quasi-WR star HD 45166 is currently one of the most
promising observational counterparts of relatively low-mass helium stars on the
helium main sequence (Sect.~\ref{sect:counterpart}).  Its mass is about
4.2~\Msun{} with a surface helium mass fraction of 0.5 and surface luminosity
of $\log L/\mathrm{L_\odot} = 3.75$. This star is likely on the helium main
sequence. The estimated mass loss rate gives a better agreement with the
extrapolated value of the Nugis \& Lamers rate than the mass loss rates of
extreme helium stars of \citet{Hamann82}.  But we still have only one sample
for such relatively low-mass helium stars in the core helium burning phase, and
cannot make a solid conclusion on which mass loss prescription is best suited
for our purpose. As shown in Fig.~\ref{fig:finalmass2}, however,  
this uncertainty does not make a great difference in terms of the final masses. 

The predicted final masses of SN Ib/c progenitors that undergo Case B/BB mass
transfer are given by Fig.~\ref{fig:finalmass2}.   The models in the
figure, for which the considered initial period of the orbit ranges from 4 to 7
days, were taken from \citet{Wellstein99} and \citet{Yoon10}.  In
\citet{Wellstein99},  conservative mass transfer (i.e., the mass accretion
efficiency  $\beta$, which is the ratio of the transferred matter from the
primary to the accreted matter onto the secondary, equals to 1.0) was assumed
and rotation was not taken into account.  \citet{Yoon10} considered the effects
of rotation, with which $\beta$ is self-regulated  by the interplay between the
mass transfer from the primary star and the mass-loss enhancement due to
rotation from the secondary star that is spun-up by mass and angular momentum
accretion.  These models indicate $\beta \simeq 0.7$ for Case A mass transfer
and $\beta = 0.0 \sim 0.8$ for Case B mass transfer, respectively.

For Case B systems, the final masses converge to about 3.15~\Msun{}  with
$f_\mathrm{w} = 1.0$ and gradually increases with increasing
$M_\mathrm{ZAMS}$ for $f_\mathrm{w}=$~5 and 10.  The result with the Nugis \&
Lamers rate is comparable to that with $f_\mathrm{w} = 10$. 
For $M_\mathrm{ZAMS} \lesssim 18$~\Msun{}, helium stars produced by Case B mass
transfer undergoes Case BB mass transfer during carbon burning and later
phases. Consequently the final masses for Case BB systems decrease more rapidly
from this point with decreasing $M_\mathrm{ZAMS}$, and the lower boundary of
the ZAMS mass for SN Ib/c progenitors becomes about 12.5~\Msun{}, below which
the primary star becomes a white dwarf. 

In a Case A system, the primary star
loses  a significant fraction of the initial mass on the main sequence, and the
helium core mass at the end of core hydrogen burning becomes lower than the
corresponding Case B system.  This makes the lower limit of ZAMS mass for SNe
Ib/c shift to about 16~\Msun{} for Case A systems \citep{Wellstein99, Yoon10}.
Very short orbits (typically $P_\mathrm{init} < 5~\mathrm{days}$) are
required for Case A systems, and their contribution to the SNe Ib/c rate from
binary systems may be smaller by about a factor of 3 -- 4 than  that of Case B
systems \citep[cf.][]{Sana12}.
 
These minimum ZAMS mass limits for SNe Ib/c (i.e., 12.5~\Msun{} and 16~\Msun{}
for Case B and Case A, respectively)  are significantly higher than for SNe IIP
from single stars, which is about 8 - 9~\Msun{}~\citep{Smartt09, Ibeling13}.
Note that binary interactions can  make this limit for SN IIP lowered even down
to about 4~\Msun{}: for example, mergers of 4~\Msun{} plus 4~\Msun{} star would
make a 8~\Msun{} SN IIP progenitor.  Therefore, the stronger association of SNe
Ib/c with younger stellar populations than SNe IIP in their host
galaxies~\citep[e.g.,][]{Anderson12, Kelly12, Sanders12} is in qualitative
agreement with the binary scenario.

According to recent analyses of SN light curves \citep{Drout11, Cano13,
Lyman14, Taddia14},  SNe Ib/c have typically ejecta masses of 1~\Msun{} --
6~\Msun{}\footnote{However, the systematic uncertainty on the inferred ejecta
masses based on SNe Ib/c light curves can be large (up to a factor of 4 in principle)
depending on the assumption for opacity and the method for measuring the light
curve width (F. Bianco and M. Modjaz, private communication)}.  This means
that, assuming the remnant mass of  1.4~\Msun{}, the final masses of ordinary
SNe Ib/c range from 2.4~\Msun{} to 7.4~\Msun{} at least. If some amounts of
helium were hidden in these analyses as argued by \citet{Piro14}, the actual
final mass would be somewhat higher.  This observation is not compatible with
the case of $f_\mathrm{w} = 1$ in Fig.~\ref{fig:finalmass2}, which predicts too
low ejecta masses on average. The results with lower mass loss rates give a
better agreement: stars of $M_\mathrm{ZAMS} =$ 14~\Msun{} - 35~\Msun{} can
explain the observed ejecta mass range with $f_\mathrm{w} = 10$ and the Nugis
\& Lamers rate for example.  \citet{Lyman14} also gives a similar conclusion
from the comparison of their stellar population model with the observation. 

The effect of different values of $f_\mathrm{w}$ 
can be regarded as the metallicity effect, because both observations and
theories indicate that WR mass loss becomes stronger for higher
metallicity~\citep{Vink05, Crowther07, Graefener08}.  The result in
Fig.~\ref{fig:finalmass2} implies that SN Ib/c ejecta should be systematically
more massive for lower metallicity, which can be tested by observations.
If there were a certain mass cut for the boundary between  successful
SN explosion and black hole formation in terms of the final mass, this
metallicity dependence of final masses would lead to gradual decrease of the SN
Ib/c rate for decreasing metallicity ~\cite[cf.][]{Boissier09, Arcavi10}. For example, if
$M_\mathrm{f} = 8$~\Msun{} were the lower limit for BH formation, we would not
expect a SN Ib/c explosion from  $M_\mathrm{ZAMS} \gtrsim 40$~\Msun{} at solar
metallicity ($f_\mathrm{w} \sim 10$) and  $M_\mathrm{ZAMS} \gtrsim 25$~\Msun{}
at SMC metallicity ($f_\mathrm{w} \sim 20$), respectively.

\subsection{Helium}\label{sect:binaryhe}

\begin{figure}
\begin{center}
\includegraphics[width=\columnwidth]{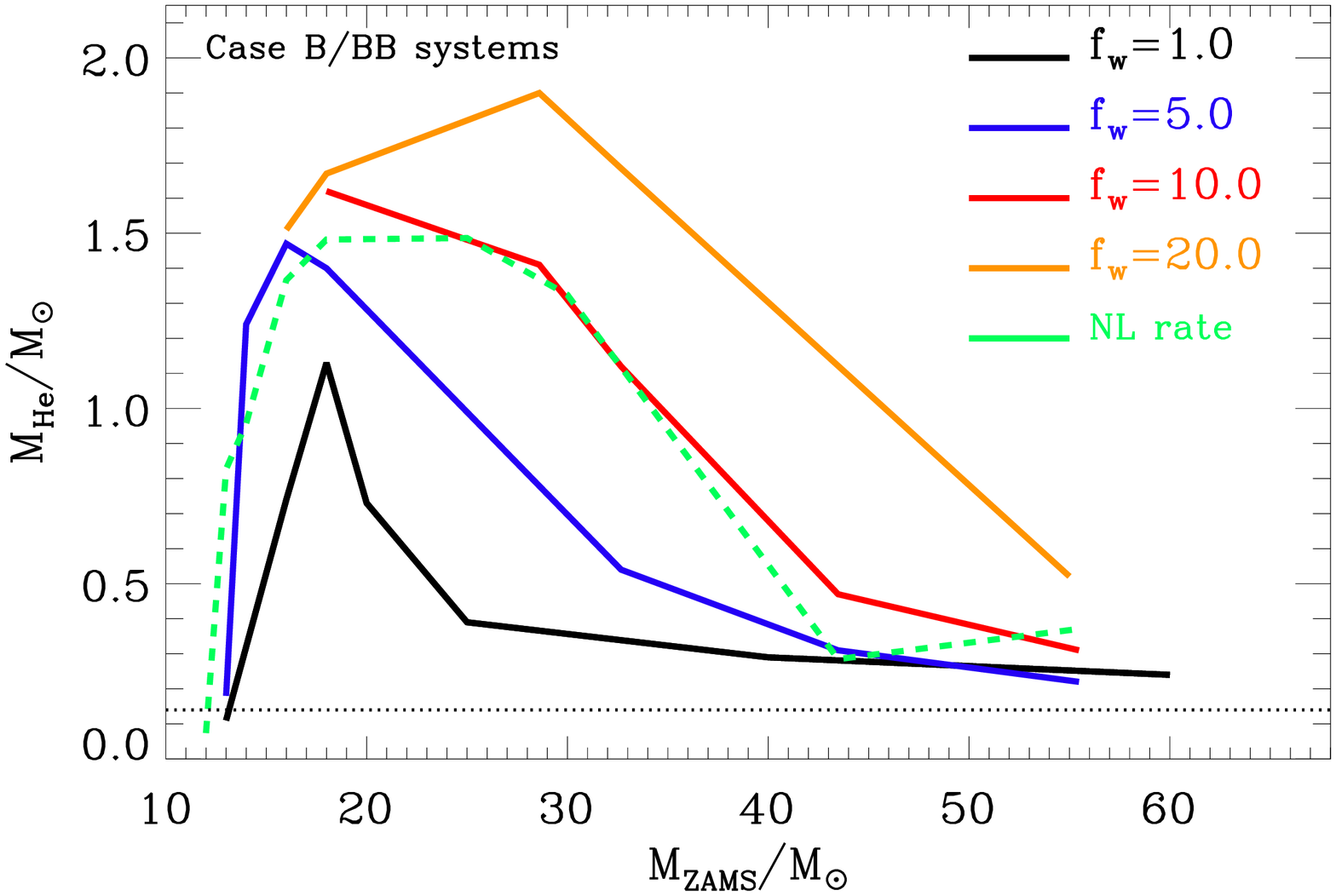}
\includegraphics[width=\columnwidth]{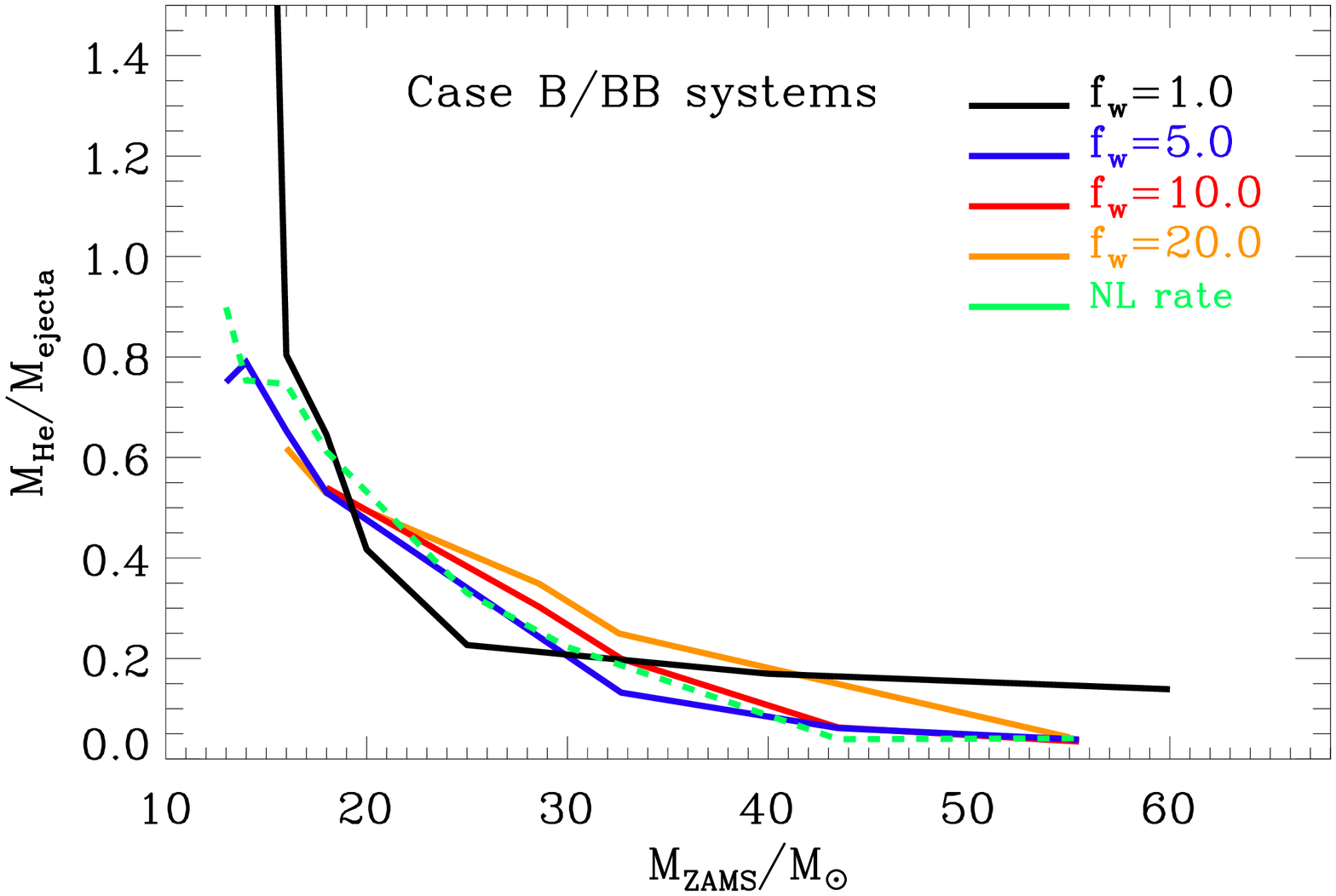}
\caption{\emph{Upper panel}: The total amounts of helium that are retained until the pre-SN stage
in SN Ib/c progenitors via Case B/BB mass transfer, as a function 
of the initial mass for different loss rates of helium stars.
Here  $f_\mathrm{w}$ denotes the reduction factor that are applied to the mass loss rate given by Eq.~(1) (see the figure caption of Fig.~\ref{fig:finalmass2}). 
The NL rate means the WR mass loss rate by \citet{Nugis00}. 
The data were  taken from  \citet{Wellstein99} for $f_\mathrm{w} = 1$, \citet{Yoon10} for $f_\mathrm{w} =$ 5, 10 and 20, 
and unpublished models by Yoon (in prep.) for the NL rate. 
\emph{Lower panel}: The corresponding ratios of the helium to ejecta mass. Here
we assumed that the remnant neutron star mass is 1.4~\Msun. See Fig.~\ref{fig:finalmass2}
for the corresponding final mass.}\label{fig:hemass}
\end{center}
\end{figure}

As discussed in Sect.~\ref{sect:singlehe}, the amount of helium retained in
progenitors may be one of the key factors that make the difference between SNe
Ib and SNe Ic.  As shown in Fig.~\ref{fig:hemass}, helium mass varies from  0.1
to  1.9 for the considered range of ZAMS mass and mass loss rates  in Case B/BB
systems.  Helium mass ($M_\mathrm{He}$) as a function of
$M_\mathrm{ZAMS}$ has a local maximum  at $M_\mathrm{\odot} \sim 17 - 18$~\Msun{}
for $f_\mathrm{w}=$~1 and 5, and  at $M_\mathrm{\odot} \sim 30$~\Msun{} for
$f_\mathrm{w}=$~20, respectively.  The rapid decrease of $M_\mathrm{He}$ as
$M_\mathrm{ZAMS}$ approaches 12~\Msun{} results from  Case BB mass transfer,
while the gradual decrease of $M_\mathrm{He}$ with increasing $M_\mathrm{ZAMS}$
is the effect of the increasing mass loss rate. 

It is important to note that, even for the case of very strong mass loss (i.e.,
$f_\mathrm{w} = 1$), the amounts of helium at the pre-SN stage are
significantly greater than the upper limit of $M_\mathrm{He} = 0.14$~\Msun{}
for SNe Ic that was suggested by \citet{Hachinger12}, except for the extreme
Case BB case at around $M_\mathrm{ZAMS} = 12.5 - 13.5$~\Msun{}. Therefore, both
single and binary star models cannot fulfill the condition of low amounts of
helium for SN Ic.   This might mean that, in reality, the chemical mixing
between helium and nickel in SN ejecta is not as efficient as considered by
\citet{Hachinger12} for most SNe Ic \citep[cf.][]{Dessart12} such that more
helium could be hidden. Such inefficient mixing is not usually supported by
observations \citep[e.g.,][]{Hachinger12, Cano14, Taddia14} \footnote{Liu, 
Modjaz et al. (2015, private communication) also show that the observed photospheric velocities and
the equivalent widths of the O I 7774 line of SNe Ib/c are not compatible with
the predictions from progenitor models with a large amount of unmixed helium}, 
but we need a systematic study on how the mixing efficiency via the
Rayleigh-Taylor instability depends on the pre-SN structure of SNe Ib/c
progenitors to resolve this issue.
Mass loss from helium stars 
is another uncertain factor that should be better understood. 
In particular, some helium stars can closely approach the Eddington limit
during the final evolutionary stages if the mass is high enough.  This might
make the surface layers unstable to cause rapid mass
eruption~\citep[cf.,][]{Maeder12, Graefener13}, thus removing most of helium in
the envelope shortly before the SN explosion as evidenced by many SNe
Ib/c~\citep{Foley07, Wellons12, Gorbikov14, GalYam14} . 

\citet{Eldridge11} assumed  a certain value of the ratio of the helium mass to
the ejecta mass ($M_\mathrm{He}/M_\mathrm{ejecta}$) as the demarcation
criterion between SN Ib and SN Ic. Because the chemical mixing between helium
and nickel plays an important role for having helium lines in SN
spectra, using $M_\mathrm{He}/M_\mathrm{ejecta}$ instead of $M_\mathrm{He}$ may
be appropriate because a lower value of  $M_\mathrm{He}/M_\mathrm{ejecta}$
means that helium can be more easily shielded from the gamma-ray photons
produced in the innermost nickel-rich layer of the SN ejecta.
Interestingly,  Fig.~\ref{fig:hemass} indicates that
$M_\mathrm{He}/M_\mathrm{ejecta}$ does not depend on the adopted mass loss rate
(hence on  metallicity) as strongly as $M_\mathrm{He}$ does. This is because,
for a lower mass loss rate, the final mass and the CO core mass become higher
and compensate the higher helium mass.  This has important
consequences in the prediction of SN Ib/c rate as a function of metallicity as
discussed in Sect.~\ref{sect:fate} below.  

If  relatively low-mass helium star progenitors in binary systems have helium
masses of about 1 - 1.5~\Msun{} in their envelopes as the most recent models
predict,  this may lead to an early plateau phase due to helium recombination
for several days as shown by \citet{Dessart11}.  This prediction has recently
been tentatively confirmed with the early-time light curve of SN Ib
2006lc~\citep{Taddia14}.  Even in the absence of non-thermal effects, helium
lines will also be observed during this phase, given that the mass fraction of
helium in the envelope is very high ($\sim 0.98$; see
Sect.~\ref{sect:singlehe}).  

On the other hand, SN Ib/c progenitors with a compact star companion (see
Fig.~\ref{fig:binary}) may undergo very strong mass transfer after core helium
exhaustion, because these systems have a small mass ratio and short orbital
period. Most of the helium envelope can be stripped off in this case, and very
little helium will be left at the pre-SN stage~\citep{Pols02, Dewi02,
Ivanova03}. This evolutionary channel has been often invoked to explain SNe Ic
having low ejecta masses and fast declining light curves~\citep[e.g.,][]{Nomoto94},
including some ultra-faint SNe Ic~\citep{Tauris13}.  The frequency of such
events would be rare compared to ordinary SNe Ib/c, and its quantitative
prediction heavily depends on the uncertain parameters related to the common
envelope efficiency and neutron star kick.

\subsection{Hydrogen}

Case B/AB mass transfer in a binary system stops when the hydrogen envelope of
the primary star is almost stripped off, but it does not completely remove
hydrogen. The amount of hydrogen remaining in the outermost layers of the
helium core immediately after the mass transfer phase is typically $0.05 -
0.1$~\Msun{}.  \citet{Yoon10} found that the final amount of hydrogen at the
pre-SN stage is a function of the progenitor mass as the following.  At solar
metallicity, which roughly corresponds to the case with $f_\mathrm{w} = 5
\cdots 10$ in Figs.~\ref{fig:finalmass2} and~\ref{fig:hemass} (see also
Fig.~\ref{fig:wrwind}), no hydrogen will be left for helium stars having final
masses of $M_\mathrm{f} \gtrsim 4 \cdots 4.4$~\Msun{}.  For the case of
$M_\mathrm{f} \lesssim 3$~\Msun{}, Case BB/ABB mass transfer becomes efficient
enough to completely remove hydrogen.  For the final mass range in-between, the
remaining hydrogen until SN explosion will be about $10^{-4} \cdots
10^{-2}$~\Msun{}.  On the other hand, the models with the Nugis and Lamers mass
loss rate at solar metallicity lose all the hydrogen in the envelope for the
whole mass range~(\citealt{Eldridge13}; Yoon, S.-C. in prep).  Therefore, in
principle, the signature of hydrogen in SN Ib/c progenitors can indirectly
constrain the mass loss rate from helium stars if we have good information
about the metallicity in the vicinity of the SN site. 

This small amount of hydrogen can be easily detected in the early time spectra
of the resulting SN \citep{Spencer10, Dessart11}, in which case it will be
identified as a SN IIb ~\citep[cf.][]{Chornock11}.  These progenitors have
relatively small radii ($R = 8 \cdots 50~\mathrm{R_\odot}$) compared to the
yellow-supergiant SN IIb progenitors~\citep[e.g.][]{Maund04, VanDyk14} produced
via Case C mass transfer~\citep{Podsiadlowski93, Claeys11}, and may belong to
the compact category of SN IIb \citep{Chevalier10}.   Some authors argue that
several SNe classified as Type Ib also have weak hydrogen absorption lines at
high velocity~\citep{Deng00, Branch02, Elmhamdi06}.  Therefore, whether the
explosion of a helium star having a thin hydrogen layer may be recognized as SN
IIb or SN Ib  depends on the details of the SN observation. 

The progenitor mass range for which a thin hydrogen layer is present at the
pre-SN stage becomes widened and the total amount of remaining hydrogen
increases with decreasing metallicity. The ratio of SN Ib/c to SN IIb
rate should decrease with decreasing metallicity (Fig.~\ref{fig:fate}).

\subsection{Supernova Types}\label{sect:fate}

\begin{figure}
\begin{center}
\includegraphics[width=0.7\columnwidth, angle=-90]{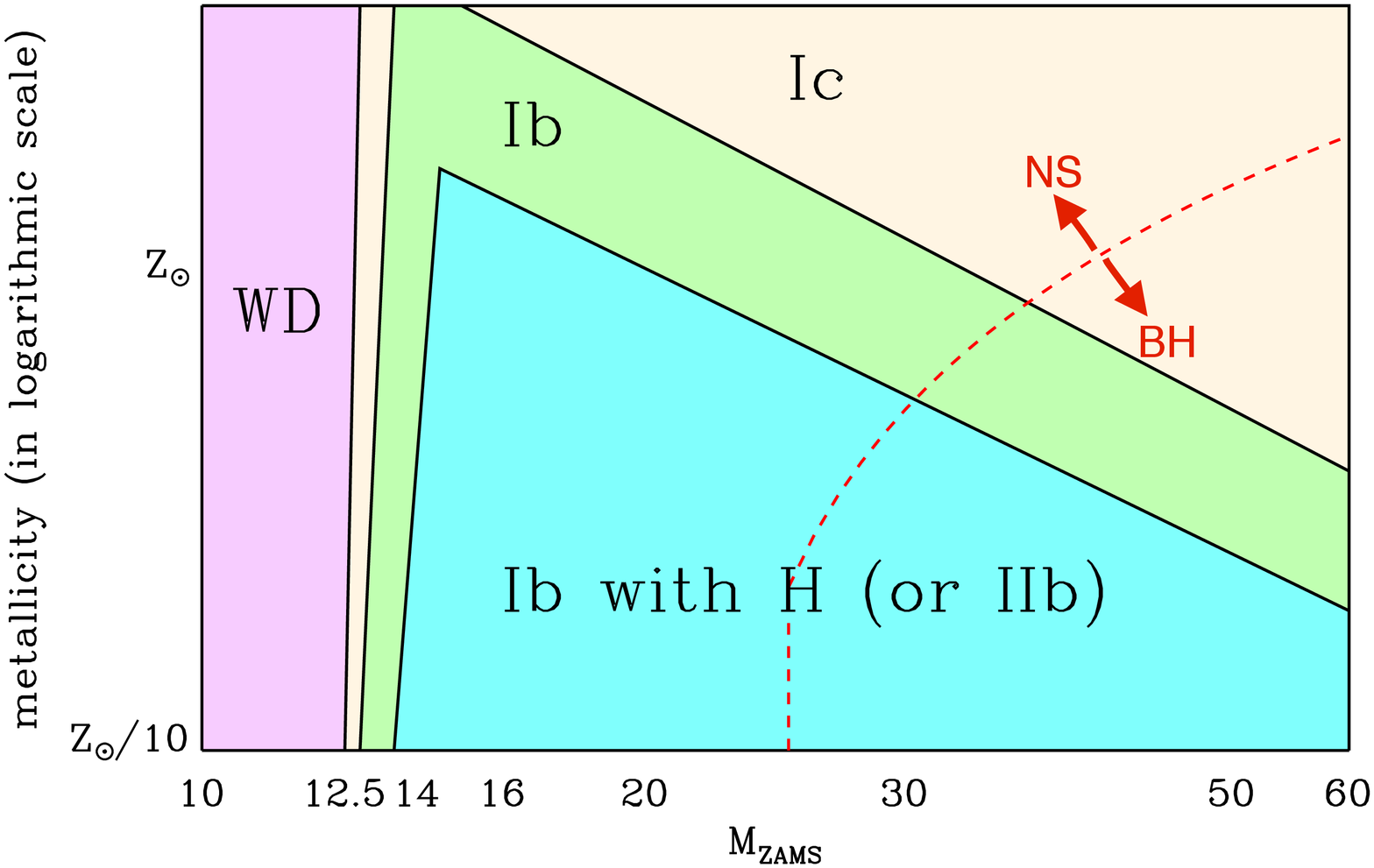}
\includegraphics[width=0.7\columnwidth, angle=-90]{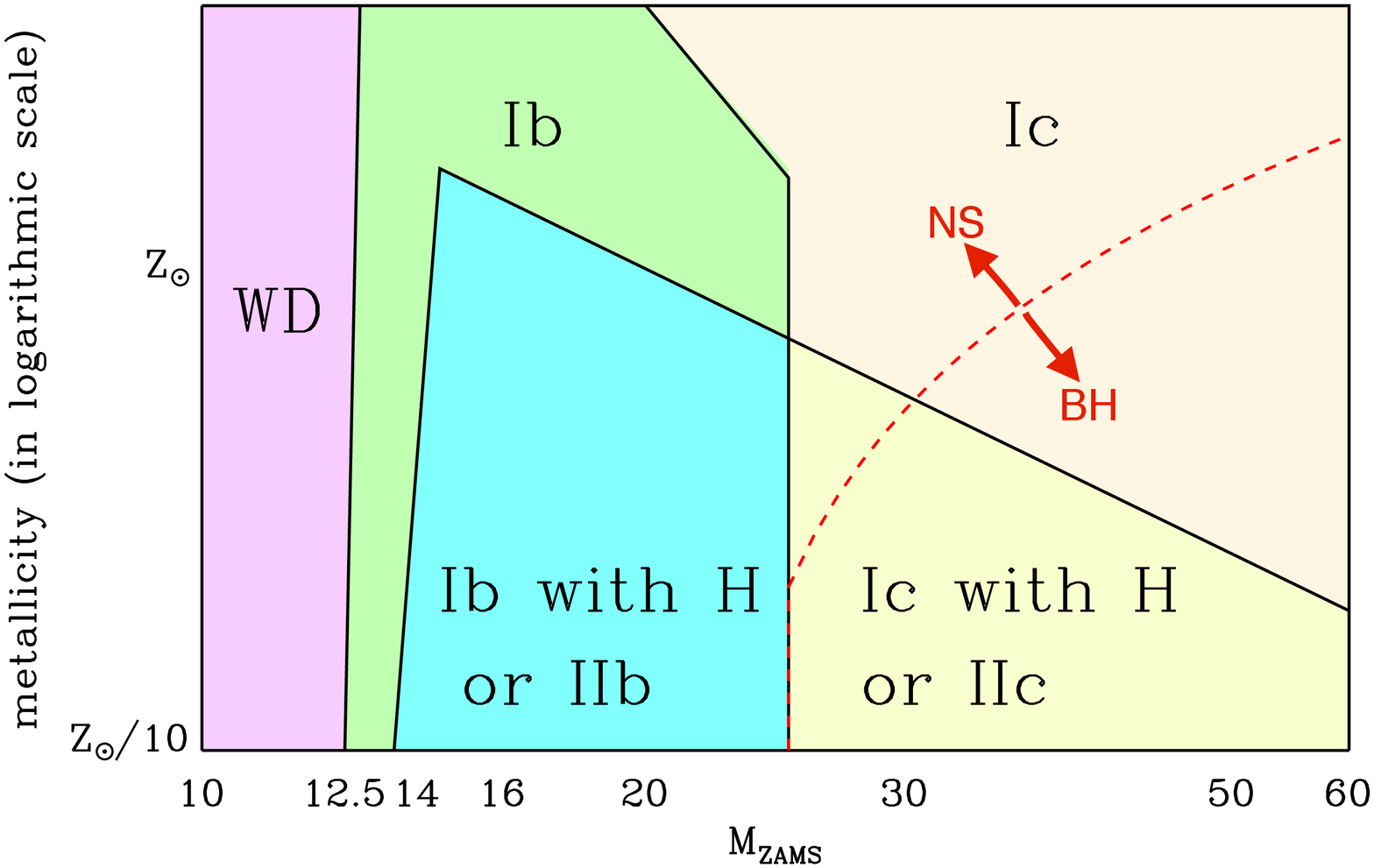}
\caption{The predicted supernova types according to the initial mass and metallicity of primary stars in Case B/BB binary systems, 
based on the result presented in Figs.~\ref{fig:finalmass2} and~\ref{fig:hemass}. 
Here $M_\mathrm{He} = 0.5$ and $M_\mathrm{He}/M_\mathrm{ejecta} = 0.45$  
are adopted for the demarcation condition between SN Ib and SN Ic, for the upper (CASE I) and  lower (CASE II) panels, respectively. 
The red dashed line denotes the critical limit for BH formation, assuming that  $M_\mathrm{f} > 8.0$~\Msun{} does not 
results in a neutron star (NS) remnant. Note that the figure provides only a qualitative prediction and 
the numbers that determine each boundary are subject to significant modification depending on the adopted
assumptions.
}\label{fig:fate}
\end{center}
\end{figure}

From the above discussion, we can make a crude prediction on the SN 
types from Case B/BB mass systems as summarized in Fig.~\ref{fig:fate}.  With
Case AB/ABB systems, each boundary in the figure would simply shift to a higher
$M_\mathrm{ZAMS}$.  For example, the lower limit of $M_\mathrm{ZAMS}$ for SN
Ib/c from Case B/BB systems is about 12.5~\Msun{}, while it is about 16~\Msun{}
for Case AB/ABB systems.  Because we still do not have any clear demarcation
between SN Ib and SN Ic in terms of the progenitor structure
(Sect.~\ref{sect:binaryhe}), ad-hoc assumptions of $M_\mathrm{He} = 0.5$~\Msun{} (CASE
I) and  $M_\mathrm{He}/M_\mathrm{ejecta} = 0.45$ (CASE II) were made for the
upper and lower panels, respectively. 

Several interesting predictions can be made from this figure, which should be tested in future observations. 
These predictions are only relevant for ordinary SNe Ib/c and those associated with GRBs or SLSNe-Ic are not considered here. 
\begin{itemize}
\item For CASE I,  the ratio of SN Ic to SN Ib rate from binary systems would  
decrease with decreasing metallicity in good agreement with \citet{Arcavi10} and \citet{Modjaz11}, and 
SN Ic is hardly expected at sufficiently low metallicity. 
By contrast, this ratio would not necessarily decrease for decreasing metallicity with CASE II, 
unless there existed a mass cut of $M_\mathrm{ZAMS}$ for BH formation ($M_\mathrm{cut}$). 
If SN Ib/c progenitors with a sufficiently high final mass could not produce an ordinary SN Ib/c, 
the SN Ic rate would significantly decrease with decreasing metallicity even with CASE II.
\item For CASE II, SNe Ic with hydrogen or SNe IIc are expected to occur at sufficiently low metallicity~\citep[cf.][]{Elmhamdi06}. 
 \citet{Dessart12} indeed showed that SNe IIc can be produced if helium  is effectively shielded from radioactive nickel.
\item The SN Ib/SN IIb ratio would decrease with decreasing metallicity for both cases, which is in good agreement with the recent observation by \citet{Arcavi10}. 
\item  Comparison of Fig.~\ref{fig:finalmass2} and Fig.~\ref{fig:fate} implies that both the initial and  final masses of SNe Ic progenitors would be  systematically higher than 
those of SN Ib progenitors, regardless of the adopted demarcation criterion of helium. 
This prediction is in qualitative agreement with SN observations~\citep{Cano13, Lyman14, Taddia14} 
and the stronger association of SNe Ic with younger stellar population than SNe Ib in the host galaxies~\citep[e.g.,][]{Anderson12, Kelly12, Sanders12}. 
\item For CASE II, the average ejecta masses would increase following the order of SN Ib, IIb and Ic at sufficiently low metallicity. 
\item The average ejecta masses of SN Ib/c would increase with decreasing metallicity. This effect would be more dramatic for SNe Ic than
for SNe Ib (see Fig.~\ref{fig:finalmass2}). 
\item The comparison of Fig.~\ref{fig:finalmass2} and Fig.~\ref{fig:fate} indicates that for CASE II, the maximum final mass of SN Ib (and IIb) would be limited to a fairly small 
value (about 7.0~\Msun{} with the assumed value of $M_\mathrm{He}/M_\mathrm{ejecta} = 0.45$) even for very low metallicity. 
\end{itemize}

\subsection{Surface Properties and the Progenitor Candidate of iPTF13bvn}\label{sect:binarysurface}

As explained in Sect.~\ref{sect:binaryexample}, a relatively low-mass helium
star progenitor of SN Ib/c  in binary systems undergo rapid expansion of its
envelope during the carbon burning phase and later stages. This envelope
expansion  becomes stronger for a lower mass star for which the carbon-oxygen
core becomes more compact, following the  mirror effect \citep{Yoon10, Yoon12, 
Eldridge13}.  Within the framework discussed in Sect.~\ref{sect:fate},
therefore, SNe Ib progenitors would have a more extended envelope than SN Ic
progenitors for most cases, which may in turn lead a more luminous early plateau
in the consequent SN \citep{Dessart11}. 

The extended envelope at the pre-SN stage can make a SN Ib/c progenitor
fairly bright in the optical.  The expected visual magnitude is about $-4
\cdots -5$  for the progenitors having final masses of 3 -- 5~\Msun{}
(\citealt{Yoon12, Bersten14, Eldridge15, Kim15}).  
With an O-type star companion, the visual brightness would be
even higher ($M_\mathrm{V} \simeq -6 \cdots -7$).  By contrast, WR star
progenitors ($M_\mathrm{ZAMS} \gtrsim 30$~\Msun{}) from both single
and binary stars have much higher bolometric luminosities, but the expected
high surface temperatures at the pre-SN stage result in fainter visual
brightness \citep[i.e., $M_\mathrm{V} \simeq -3$ ][]{Yoon12, Groh13a} in most
cases, which would make them more difficult to directly identify in
pre-SN images as discussed in Sect.~\ref{sect:singlesurface}.   

Recently, \citet{Cao13} have reported the tentative identification of the
progenitor of the SN Ib iPTF13bvn.  The estimated absolute magnitudes of the object in
the optical  range from -5.0 to -7, depending on the filters,
adopted extinction values, and photometry methods \citep{Cao13, Bersten14,
Eldridge15}.  \citet{Groh13b} argued for a single star progenitor with initial
masses of 31 - 35~\Msun{} based on the non-rotating models of the Geneva group.
Unlike more massive stars that become WO stars at the end, these models end
their lives  as WN stars having relatively thick helium envelopes with surface
temperatures of about $\log T_\mathrm{eff} \sim 4.6$. The predicted optical
magnitudes ($\sim -5.5$) agree with the observation, but the final mass ($\sim
11$~\Msun) seems to be too high, compared to the estimated ejecta mass of the
SN ($\sim 1.9 - 2.3$~\Msun{}; \citealt{Fremling14, Bersten14}) that
supports the binary scenario. The optical brightness of the progenitor candidate
can also be explained by a binary progenitor having an initial mass of 10 -
20~\Msun{}~\citep{Bersten14, Eldridge15}. If future observations find evidence
for the companion star that survives the SN explosion, it will directly
confirm a binary star origin of SNe Ib for the first time.  

\subsection{Rotation}\label{sect:binaryrot}

Massive stars in a close binary system before mass transfer are likely to be synchronized with the
orbit because of the short tidal interaction timescale.  After Case B/AB mass
transfer, the orbit becomes too wide to  keep the tidal synchronization, and the final
rotation velocity of the primary star is mainly determined by the amount of
angular momentum that is retained after the mass transfer phase.  Binary
evolution models including the effects of rotation  indicate that SN Ib/c
progenitors via Case B/AB mass transfers are slow rotators as they lose mass
and angular momentum  \citep{Wellstein01a, Yoon10, Langer12}.  Both models with and without
magnetic torques due to the Tayler-Spruit dynamo (see Sect.~\ref{sect:singlerot} for the discussion on
angular momentum redistribution inside stars) predict that the surface rotation
velocity of naked helium stars on the helium main sequence after the mass
transfer phase is only about $1.0 - 10.0 ~\mathrm{km~s^{-1}}$.  

The specific angular momentum in the innermost region of 1.4~\Msun{} that would
collapse to a neutron star significantly according to the prescription
of angular momentum transfer \citep{Yoon10}.  Models without the Tayler-Spruit dynamo predict that
neutron stars from SNe Ib/c in binary systems would almost reach the critical
rotation.  This means that the majority of SN Ib/c progenitors in binary systems would also be
good  progenitor candidates for magnetars/collapsars without the Talyer-Spruit dynamo, which cannot
be easily reconciled with observations.  Models with the Tayler-Spruit dynamo predict that the
specific angular momentum  of the collapsing core would be comparable to those
of single star models (i.e., $j \sim 10^{14}~\mathrm{cm~s^{-1}}$) and the
resultant neutron stars would be rotating at a period of several milliseconds.

 As in the case of single stars, therefore, binary models with the Tayler-Spruit
dynamo do not predict magentar/collapsar progenitors for energetic SNe and/or
long GRBs  via the standard Case B/BB/AB/ABB systems~\citep{Yoon10} under
normal circumstances.  At sufficiently low metallicity, however, mass accreting
stars in Case B mass transfer systems may be spun up to undergo the chemically
homogeneous evolution, which may end up as a GRB~\citep{Cantiello07}.  SN Ib/c
progenitors in very close binary systems consisting of a helium star and a
low-mass main sequence star/compact object  (cf. Fig.~\ref{fig:binary};
Sect.~\ref{sect:binaryexample}) may experience strong tidal interaction, given
that the orbital period can be as short as 0.1~day in this case
\citep[e.g.][]{Dewi02, Ivanova03, Izzard04, Heuvel07, Detmers08, Podsiadlowski10}.  Some of
these systems may produce unusually rapid rotators even within the framework of
the Tayler-Spruit dynamo, but more detailed evolutionary studies  are needed to
make a meaningful conclusion on this.  The fact that host galaxies of
broad-lined SNe Ic and GRBs are found to be systematically overdense compared
to other SDSS galaxies might be an indication for the importance of the binary
channel for producing rapidly rotating progenitors to produce these
events~\citep{Kelly14}.

\subsection{Observational Counterparts}\label{sect:counterpart}

WR stars in  binary systems are excellent observational counterparts for SN Ib/c
progenitors during the post Case B/AB mass transfer phase, for $\log
L/\mathrm{L_\odot} \gtrsim 5.0$~\citep{Hucht01}.  Relatively low-mass helium stars ($\log
L/\mathrm{L_\odot} < 5.0$) in binary systems are only rarely observed.  This is
probably because such helium stars on the helium main sequence are very hot and
faint in optical bands (see Fig.~\ref{fig:hr2}) and because many of them have
bright OB-type companion stars.  

As mentioned above (Sect.~\ref{sect:binarymass}), the best observational
counterpart of binary SN Ib/c progenitors with $\log L/\mathrm{L_\odot} < 4.5$
is the quasi-WR (qWR) star HD 45166 \citep{vanBlerkom78, Willis83, Steiner05,
Groh08}: the primary is a helium rich 4.2~\Msun{} star with $R \simeq
1.0~\mathrm{R_\odot}$ and the secondary is a 4.8~\Msun{} main sequence star, in
a 1.596 day orbit \citep{Steiner05}.  This system was probably produced via
unstable Case B/AB mass transfer and the consequent common envelope phase.
This star gives important information about the mass loss rate from such a
relatively low-mass helium star as discussed in Sect.~\ref{sect:binarymass}.
WR 7a is another qWR star~\citep{Scwartz90, Pereira98, Oliveira03}.  No
companion of this star has been found so far.  Radial periodicity of 0.204 day
has been reported, and if this is related to binarity, its companion should be
a low-mass main sequence star ($M\lesssim 1.0$~\Msun) or a compact object. To
our knowledge, no estimate of the wind mass loss rate from this star has been
reported yet.

There also exist candidates for evolved helium giant stars beyond core helium
exhaustion.  They include $\upsilon$ Sgr, KS Per and LSS 4300~\citep{Dudley93}.
Among these, the $\upsilon$ Sgr system has been best
studied~\citep[e.g.,][]{Frame95, Saio95, Koubsky06, Netolicky09, Kipper12}.
The mass, surface temperature and bolometric luminosity of the
hydrogen-deficient primary of this system are  $M \approx 3.0$~\Msun{},
$T_\mathrm{eff} \approx 11800$~K and $\log L/\mathrm{L_\odot} \approx 4.6$.  It
also shows radial pulsation of a 20 day period \citep{Saio95}, and evidence for
mass transfer and a circumbinary disk \citep{Netolicky09}, which agrees well
with the Case BB mass transfer scenario~\citep{Schoenberner83}. Therefore,
these systems can provide useful information  about the progenitor evolution as
well as their circumstellar environments immediately before SN
explosion. The visual magnitude of this star is $M_\mathrm{V} =
-4.73\pm0.3$~\citep{Kipper12}, which is consistent with the model predictions
discussed in Sect.~\ref{sect:binarysurface}.

\subsection{Companion Stars}

The companion stars of binary progenitors of SNe Ib/c will survive the
SN explosion and may be found in some young SN remnants
\citep[e.g.][]{Kochanek09, Koo11}. There may be several types of companion
stars:  main sequence stars of early to late types, compact objects like white
dwarfs, neutron stars and black holes, and helium stars
(Sect.~\ref{sect:binaryevol}; Fig.~\ref{fig:binary}). In terms of stellar population, the most common
type (i.e., more than 30\% of all binary SN Ib/c progenitors) may be
relatively high mass stars (O/B type) on the main sequence that underwent
stable Case B/BB/AB/ABB mass transfer \citep[e.g.,][]{Podsiadlowski92, Eldridge13}.
These massive companion stars accrete mass and angular momentum via stable mass
transfer to be spun up to the critical value~\citep{Wellstein01, Petrovic05b,
Cantiello07, Yoon10, Langer12, deMink13}.  At solar metallicity, however, they
will lose angular momentum again via stellar winds after the mass transfer
phase, and the rotation velocity shortly after the SN explosion depends on
how much mass is lost by winds until that time.  Models by \citet{Yoon10}
indicate that the surface rotation velocity will be about 300 -
450~$\mathrm{km~s^{-1}}$ for surviving companion of $M_2
\approx 17 - 20$~\Msun{}. By contrast, more massive stars loses mass and
angular momentum very quickly:  a  48~\Msun{} companion star  
would be slowed down to about 60~$\mathrm{km~s^{-1}}$  at the pre-SN
stage.  The transferred mass to the secondary star is enriched with the ashes
of hydrogen burning compared to the initial composition \citep{deMink09,
Langer12}.  The surface composition of the surviving secondary star after the
SN explosion should be therefore marked by the enhancement of helium and
nitrogen.  The models of \citet{Yoon10} give typically the mass fractions of
helium and nitrogen of about  0.35 and $4.3\times10^{-3}$ (i.e., 4.3 times the
solar value) at the surface of surviving companion stars, respectively. 

\section{CONCLUSIONS}

\begin{table*}
\begin{threeparttable}
\caption{Main predictions of single and binary star progenitors models for SNe Ib/c at solar metallicity}\label{tab1}
\begin{tabularx}{0.98\linewidth}
{>{\arraybackslash}X | >{\arraybackslash}X| >{\arraybackslash}X}
\hline
      &   Single     &   Binary       \\
\hline
Initial Mass &  $M_\mathrm{ZAMS} \gtrsim 25$~\Msun{} at $Z = 0.02$ \tnote{a}  &   $M_\mathrm{ZAMS} \gtrsim 12$~\Msun{} \tnote{b}  \\
Final Mass &   $ 10 \lesssim M/\mathrm{M_\odot} \lesssim 17 $ at $Z = 0.02$ \tnote{a}  &   $ 1.4 < M/\mathrm{M_\odot} \lesssim 17 $ at $Z = 0.02$ \tnote{c}   \\
Final Radius &  $0.5 \lesssim R/\mathrm{R_\odot} \lesssim 10$ \tnote{d}     &          $  0.5 \lesssim R/\mathrm{R_\odot} \lesssim 100 $  \tnote{e} \\
Wind Mass Loss Rate  &   $\dot{M} = 10^{-6}  \sim  5\times 10^{-5}~\mathrm{M_\odot~yr^{-1}}$ &   $\dot{M} = 10^{-7} \sim  5\times10^{-5}~\mathrm{M_\odot~yr^{-1}}$ \\
Escape Velocity ($\approx v_\mathrm{wind}$ )    &  $v_\mathrm{esc}  = 500  - 2500~\mathrm{km~s^{-1}}$   &   $ v_\mathrm{esc}  = 60 -  2500~\mathrm{km~s^{-1}}$   \\
Circumstellar Structure & $\rho \propto r^{-2}$                    & Complex  with wind-wind collision and orbital motion \\
Optical Magnitudes  & $M_\mathrm{V} \approx -3$ for WO type progenitor, and $M_\mathrm{V} =  -5.5 \cdots -6.5$ for  WN type progenitor. \tnote{f} 
                    & $M_\mathrm{V} \approx -3 \cdots -6.0$ for most helium star progenitors. It will be more luminous in optical bands with a bright companion. \tnote{g} \\ 
Light Curves &   Relatively faint at early times. Broad light curves   & Fairly luminous early time plateau for $ 13 \lesssim M_\mathrm{ZAMS} \lesssim 25$~\Msun  \tnote{h}\\
Spectra      &   No or weak helium lines for most cases, and hence biased towards the production of SNe Ic \tnote{h}.   & Helium lines even without non-thermal processes during  the early time plateau phase for   $13 \lesssim M_\mathrm{ZAMS} \lesssim 25$~\Msun{}  \tnote{h} \\
\hline
\end{tabularx}
\begin{tablenotes}
 \item[a] \citet{Meynet03}. 
 \item[b] \citet{Wellstein99, Yoon10, Eldridge13}
 \item[c] Based on the result presented in Fig.~\ref{fig:finalmass2} and \citet{Yoon10} (In the figure, $f_\mathrm{w} = 10$  roughly  corresponds to $Z = 0.02$). See also \citet{Eldridge11}. 
 \item[d] \citet{Groh13a, Groh13b}
 \item[e] \citet{Yoon10, Yoon12, Eldridge13, Eldridge15}. 
 \item[f] \citet{Yoon12}, \citet{Groh13a, Groh13b}
 \item[g] \citet{Yoon12, Bersten14, Eldridge15, Kim15}
 \item[h] \citet{Dessart11, Dessart12}
\end{tablenotes}
\end{threeparttable}
\end{table*}

We summarize the main predictions of single and binary star models for ordinary
SN Ib/c progenitors in Table~\ref{tab1}.  

Many of WR progenitors would end their lives as a WO type star that is
relatively faint with optical filters ($M_\mathrm{V} \approx -3$;
\citealt{Yoon12, Groh13a}).  Binary systems with a sufficiently high initial
mass of the primary star can produce WR progenitors of which the properties
would be very similar to those of single WR stars, but the presence of the
companion star would result in rather complex structures of the circumstellar
medium and the SN remnant~\citep{Koo11}.  Compared to the single star case, the
detectability of WR progenitors can be significantly enhanced with a luminous
early type companion.

However, given the preference for lower masses of the initial mass function,
the majority of binary progenitors should have several unique properties that
are very different from WR progenitors.  Their final masses at the
pre-SN stage are systematically lower ($ M_\mathrm{f} \simeq 1.4
- 6~\mathrm{M_\odot}$ for $M_\mathrm{ZAMS} \simeq 12 - 25~\mathrm{M_\odot}$)
  than  those of WR progenitors ($M_\mathrm{f} > 10$~\Msun{}). This agrees well
with the recent ejecta mass estimates of ordinary SNe Ib/c  ($M_\mathrm{ejecta}
= 1 - 6$~\Msun{}; \citealt{Drout11, Cano13, Lyman14, Taddia14}).  The current
binary model predictions are also  consistent with the observational facts that
the association of the SN locations in their host galaxies with young
stellar populations becomes stronger following the order of SN II, SN Ib and SN
Ic~(Sects.~\ref{sect:binarymass} and~\ref{sect:fate}), which is usually
interpreted within the single star scenario in the
literature~\citep[e.g.,][]{Anderson12, Kelly12, Sanders12}. 

Despite their relatively low masses, the detectability of binary progenitors in
optical bands is not necessarily  lower than single WR star progenitors.  To
the contrary, a significant fraction of binary progenitors should have very
high visual luminosities because relatively low-mass helium stars  can rapidly
expand during the late evolutionary stages, and/or because many of them would
have luminous companion stars \citep{Yoon12, Eldridge13, Eldridge15}.  

The SN models by \citet{Dessart11} indicate that early time light curves and
spectra would have the critical information about the nature of SN Ib/c
progenitors. In particular, the plateau phase due to helium recombination at
early times  and He I lines formed with thermal processes during this phase
would be strong evidence for a relatively low-mass helium star progenitor
having an extended helium envelope.  As demonstrated by several recent
observations\citep[e.g.,][]{Soderberg08, Modjaz09, Corsi12, Maeda14}, the
progenitor size  can also be best constrained by early-time SN light curves
including shock breakouts.  We therefore conclude that observations of
early-time light curves and spectra will be an excellent probe into the nature
of SN Ib/c progenitors.

There still exist many unsolved problems and related future topics that should
be addressed. 
\begin{itemize}
\item A caveat in the above conclusions is that the predictions summarized in
Table~\ref{tab1} are mostly based on the models at solar  metallicity.  The
final masses of single star progenitors at super-solar metallicity can become
as low as 5~\Msun{}, depending on the adopted mass loss rate
\citep[Fig.~\ref{fig:finalmass};][]{Meynet05}.  This value is within the ejecta
mass range of ordinary SNe Ib/c  by \citet{Cano13} and \citet{Lyman14},
although it still cannot explain SNe Ib/c having ejecta masses lower than about
3.5~\Msun{}.  Given that the final amounts of helium must also be smaller
than in the case of solar metallicity (Fig.~\ref{fig:hemass}),  the
contribution from super-solar metallicity single stars might be particularly
significant for SNe Ic~\citep[cf.,][]{Prieto08, Boissier09, Modjaz11}.  
\item  Although the SN Ib/c event rate  implies the dominant role of binary systems for the production of SNe Ib/c~\citep{Smith11, Eldridge13}, 
more direct evidence for binary progenitors comes from SNe Ib/c ejecta masses ($M_\mathrm{ejecta}$; Sects.~\ref{sect:binarymass} and~\ref{sect:fate}). 
Precise estimates of SN ejecta masses can give one of the best constraints for progenitor models. 
The uncertainties related to the effects of asymmetry of the explosion, mixing of chemical compositions in the SN ejecta
and  the presence of helium on the estimates of $M_\mathrm{ejecta}$ using SN light curves
should be clarified in the near future \citep[cf.][]{Dessart12, Piro14, Wheeler14}.
\item The question on  how much helium can be hidden in SN Ic spectra is another 
critical test case for SN Ib/c progenitor models (Sects.~\ref{sect:singlehe}, \ref{sect:binaryhe} and \ref{sect:fate}; Fig.~\ref{fig:fate}).  
Recent observations indicate very low helium mass  ($ < 0.14~\mathrm{M_\odot}$) in the ejecta \citep{Hachinger12, Taddia14}. This 
cannot be easily accommodated to the current model predictions that most SN Ib/c progenitors have $M_\mathrm{He} > 0.2$. 
This conflict may be due to our lack of proper knowledge on the mass loss rate from helium stars (in particular during
the post-helium burning phase; Sect.~\ref{sect:binaryhe}) but a rigorous estimate of helium masses
for a large sample of SNe Ib/c is highly required to resolve this issue.  The ratio of SN Ic to SN Ib is
another important constraint for progenitor models, which 
should be better estimated in the future. Several authors 
reported that the Ic to Ib ratio is about 2 \citep[e.g.,][]{Smartt09, Li11}, but  \citet{Modjaz14} point out 
that many SNe Ib had been misclassified as SNe Ic, suggesting a lower value.
\item  As discussed in Sect.~\ref{sect:fate}, future observations on SN Ib/c and SN IIb properties 
and their relative frequencies as a function of metallicity
\citep[e.g.,][]{Prieto08, Boissier09, Modjaz11, Graham13}
would  greatly help to clarify the role of mass loss from helium stars in the evolution of SN Ib/c progenitors. 
It would be particularly important to investigate how the ejecta masses of SNe Ib/c and the ratio of SN Ib/SN Ic 
rate systematically depend on metallicity.  
\item  In the discussions above, we did not address progenitors of broad-lined
SNe Ic (SNe Ic-BL), simply because we do not have a good clue on what makes
them. The association of long GRBs and some SNe Ic-BL ~\citep[e.g.][]{Woosley06} implies the
importance of rapid rotation, and attempts have been made to explain long GRBs,
SNe Ic-BL and super-luminous SNe Ic within the single framework of the magnetar
scenario~\citep[e.g.,][]{Mazzali14}. Binary star models including rotation do not predict any particular parameter
space where unusually rapid rotation in the core can be realized from the
standard Case B/BB/AB/ABB channels for SNe Ib/c \citep{Petrovic05b, Yoon10}, except for
the so-called chemically homogeneous evolution induced by mass accretion at
low metallicity~\citep{Cantiello07}.  Future studies should investigate
more carefully  the evolution of some specific binary systems where the condition of
rapid rotation for magnetars can be rather easily fulfilled compared to the
standard channels, such as binary systems consisting of a helium star plus a
compact object in a very tight orbit~(Fig.~\ref{fig:binary};
Sect.~\ref{sect:binaryrot}).  
\end{itemize}

\begin{acknowledgements}
The author is grateful to Norbert Langer, Stan Woosley, Luc Dessart, Hyun-Jeong
Kim, Bon-Chul Koo, G{\"o}tz Gr{\"a}fener, Jorick Vink, Wonseok Chun, Thomas
Tauris, Rob Izzard, Alexandra Kozyreva, and John Hillier for their contribution
to this work and fruitful collaboration, and John Eldridge, Maryam Modjaz,
Philipp Podsiadlowski, Ken'ichi Nomoto, Keiichi Maeda, Selma de Mink and  Cyril
Georgy for helpful discussions.  This work was supported by Basic Science
Research (2013R1A1A2061842) program through the National Research Foundation of
Korea (NRF) and by Research Resettlement Fund for the new faculty of SNU.
\end{acknowledgements}


\end{document}